\documentclass[aps,prx,twocolumn,amsmath,amssymb,nofootinbib,superscriptaddress,floatfix,reprint,longbibliography]{revtex4-1}

\usepackage[dvips]{graphicx}
\usepackage{latexsym}
\usepackage{amsmath}
\usepackage{amsfonts}
\usepackage{amssymb}
\usepackage{bm}
\usepackage{color}
\usepackage{txfonts}
\usepackage{float}
\usepackage{braket}
\usepackage{url}
\usepackage{CJKutf8}
\usepackage{braket} 
\usepackage{svg}

\usepackage[colorlinks=true, allcolors=blue]{hyperref}
\usepackage{ulem}
\usepackage{cleveref}
\usepackage{bm,color,amsmath,amssymb,mathrsfs,latexsym,graphicx,psfrag,mathtools}

\crefname{Supplement}{Supp.\,}{Supps.\,}

\creflabelformat{Supplement}{[#2#1#3]}

\begin{document}
	\newcommand{\fig}[2]{\includegraphics[width=#1]{#2}}
	\def\gsim{~\rlap{$>$}{\lower 1.0ex\hbox{$\sim$}}}
	\setlength{\unitlength}{1mm}
	\newcommand{\bea}{\begin{equation} \begin{aligned}}
    \newcommand{\eea}{\end{aligned} \end{equation} }
    \def\abs#1{\left|{#1}\right|}          			% absolute value, \abs{x} gives |x|
    \def\bra#1{\left<{#1}\right|}					% "bra"-state
    \def\ket#1{\left|{#1}\right>}					% "ket"-state

\title {Anderson localization: a density matrix approach}

\author{Ziyue Qi}
\affiliation{Beijing National Laboratory for Condensed Matter Physics and Institute of Physics,
	Chinese Academy of Sciences, Beijing 100190, China}
\affiliation{University of Chinese Academy of Sciences, Beijing 100190, China}

 \author{Yi Zhang}
 \email{zhangyi821@shu.edu.cn}
 \affiliation{Department of Physics and Institute for Quantum Science and Technology, Shanghai University, Shanghai 200444, China}
 \affiliation{Shanghai Key Laboratory of High Temperature Superconductors, Shanghai University, Shanghai 200444, China}

\author{Mingpu Qin}
\email{qinmingpu@sjtu.edu.cn}
\affiliation{Key Laboratory of Artificial Structures and Quantum Control (Ministry of Education), School of Physics and Astronomy, Shanghai Jiao Tong University, Shanghai 200240, China}

\author{Hongming Weng}
\email{hmweng@iphy.ac.cn}
\affiliation{Beijing National Laboratory for Condensed Matter Physics and Institute of Physics,
	Chinese Academy of Sciences, Beijing 100190, China}
\affiliation{University of Chinese Academy of Sciences, Beijing 100190, China}

\author{Kun Jiang}
\email{jiangkun@iphy.ac.cn}
\affiliation{Beijing National Laboratory for Condensed Matter Physics and Institute of Physics,
	Chinese Academy of Sciences, Beijing 100190, China}
\affiliation{University of Chinese Academy of Sciences, Beijing 100190, China}

\date{\today}

\begin{abstract}
Anderson localization is a quantum phenomenon in which disorder localizes electronic wavefunctions. In this work, we propose a new approach to study Anderson localization based on the density matrix formalism. Drawing an analogy to the standard transfer matrix method, we extract the localization length from the modular density matrix in quasi-one-dimensional systems. This approach successfully captures the metal-insulator transition in the three-dimensional Anderson model and in the two-dimensional Anderson model with spin–orbit coupling. It can be also readily extended to multiorbital systems. We further generalize the formalism to interacting systems, showing that the one-dimensional spinless attractive model exhibits the expected metallic phase, consistent with previous studies. More importantly, we demonstrate the existence of a two-dimensional metallic phase in the presence of Hubbard interactions and disorder. This method offers a new perspective on Anderson localization and its interplay with interactions.
\end{abstract}
%\pacs{}
\maketitle

\section{Introduction}
In real materials, the presence of disorder or imperfections is inevitable which can strongly influence their physical properties. Depending on its strength, disorder can either serve merely as a small perturbation that slightly alters transport properties, or can play a decisive role in reshaping the ground state of a system \cite{THOULESS197493,Lee_RevModPhys.57.287,altshuler1985electron,Kramer_1993}. So it is a central ingredient in understanding the electronic, optical, and magnetic responses of condensed matter systems. An iconic example of disorder-induced phenomena is Anderson localization \cite{AL50}, introduced by P. W. Anderson in 1958 \cite{anderson_PhysRev.109.1492}. 
Anderson localization is a fundamental quantum phenomenon in which disorder prevents electronic wavefunctions from spreading through the system, leading to the absence of diffusion.

\begin{figure}[htbp]
    \centering
    \includegraphics[width=1.02\linewidth]{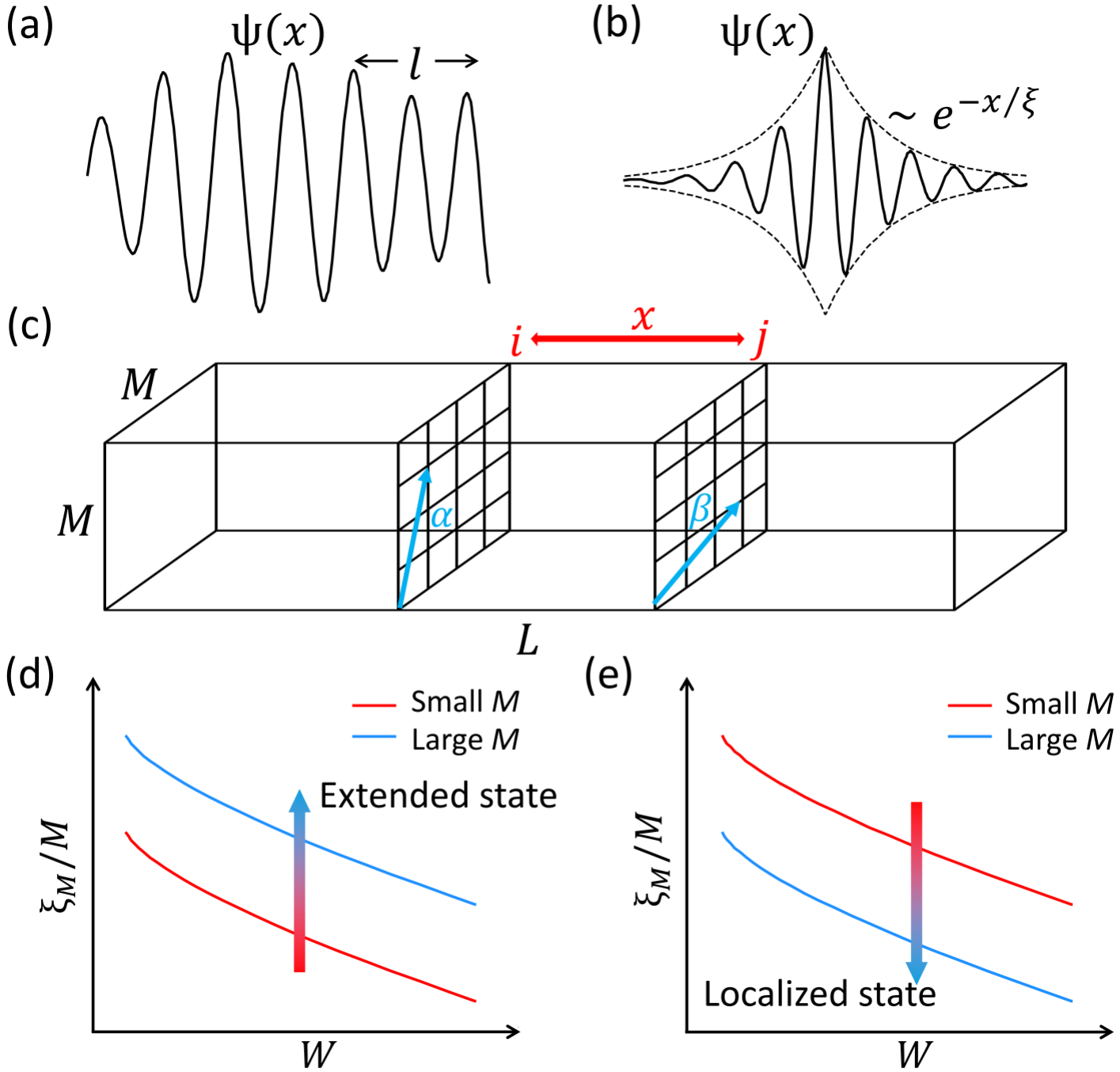}
    \caption{ Schematic diagrams of wave function localization and finite-size scaling in disordered systems. (a), (b) schematically show the real-space distribution of the extended and localized wave function respectively. The extended state exhibits undamped oscillations with finite mean free path $l$, whereas the localized state decays exponentially as $e^{-x/\xi}$. (c) shows the quasi-1D system of width $M$ and length $L$, $\alpha$ and $\beta$ are the site indices in the slice $i$ and $j$, respectively. $x$ represents the distance between slice $i$ and $j$. (d), (e) schematically show the finite-size scaling law of the dimensionaless ratio $\xi_M/M$, which increases with $M$ for the extended states and decreases with $M$ for the localized states. }
    \label{fig:schematic_diagram}
\end{figure}

The Anderson localization theory has been extremely successful over the past sixty years, providing a deep understanding of how disorder alone can localize electronic states and suppress transport \cite{Lee_RevModPhys.57.287,Kramer_1993}. A wide range of powerful methods have been developed to study this phenomenon, including scaling theory \cite{scaling_PhysRevLett.42.673,Thouless_PhysRevLett.35.1475,Wegner_scaling,Schuster_scaling},  perturbation theory \cite{altshuler1985electron,ALTSHULER1979115,Altshuler1979ContributionTT,Altshuler_PhysRevLett.44.1288}, disorder numerical methods \cite{Kramer_1993,TMM_MacKinnon_1981,p_soven_67,TMT_Dobrosavljevic_2003,tmdca_review}, and field-theoretical approaches \cite{efetov1999supersymmetry,Finkelshtein_1983}. On the other hand, the interplay between interaction and disorder remains far less understood. While interactions can fundamentally alter localization by giving rise to phenomena such as many-body localization or correlated insulating phases, a comprehensive theoretical framework that unifies disorder and interaction effects is still an open challenge in condensed matter physics \cite{MBL_RevModPhys.91.021001}.
In this work, we introduce a different perspective on Anderson localization based on the density matrix formalism. We find that this formalism is applicable to both non-interacting and interacting systems.

In general, the essential distinction between a localized state and an extended state lies in the nature of their wavefunctions \cite{THOULESS197493,Lee_RevModPhys.57.287,Kramer_1993}. In a translationally invariant metal, electronic states are well described by extended Bloch waves, which remain delocalized across the entire system, even in the presence of scattering from a random potential, with a finite mean free path $l$, as illustrated in Fig. \ref{fig:schematic_diagram}(a). In contrast, when disorder becomes sufficiently strong, the wavefunction can transition to a localized form, where its amplitude decays exponentially away from some spatial point as $e^{-x/\xi}$, as illustrated in Fig. \ref{fig:schematic_diagram}(b). Here, $\xi$ denotes the localization length, characterizing the spatial extent of the localized state.

\begin{figure*}[htbp]
    \centering
    \includegraphics[width=1.02\linewidth]{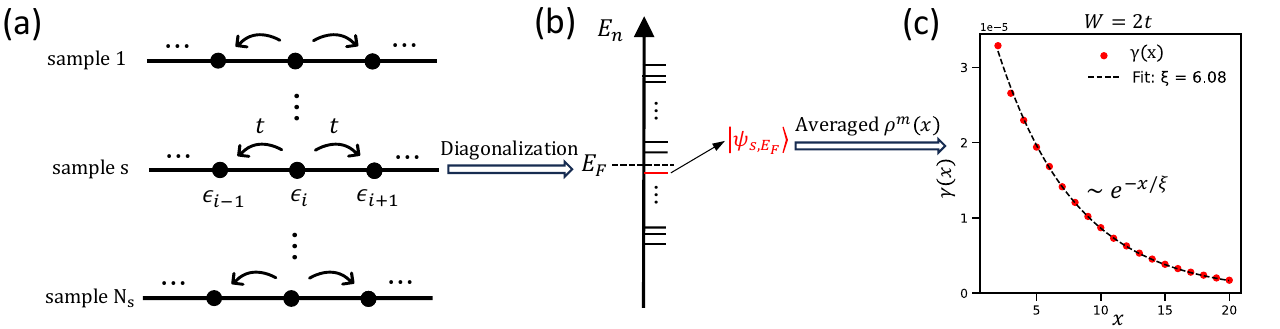}
    \caption{ Computational workflow for extracting the localization length from the MDM in one-dimensional non-interacting systems. (a) schematically shows different disorder samples with hopping energy $t$ and onsite random potential $\epsilon_i$. 
    For each disorder sample, the single-particle eigenstate with the energy level $E_n$ nearest to the Fermi energy $E_F$ is chosen to calculate $\rho^m(x)$, as illustrated in (b). (c) shows $\gamma(x)$ as a function of $x$ for the systems with disorder strength $W=2t$, which exhibits exact exponential decay with a localization length $\xi=6.08$. (c) is calculated by averaging 500 disorder realizations of length $L=120$ under periodic boundary condition at $E_F=0$.}
    \label{fig:flow_diagram}
\end{figure*}

Numerical simulations have played a crucial role in advancing the theory of Anderson localization, particularly through finite-size scaling analysis \cite{AL50}. While extracting true thermodynamic quantities directly from finite-size systems is challenging, scaling behavior offers valuable insights into localization phenomena \cite{Thouless_PhysRevLett.35.1475,Wegner_scaling,Schuster_scaling}. A prime example is the remarkable success of the single-parameter scaling approach applied to the conductance $g(L)$ of a system with linear size $L$ \cite{scaling_PhysRevLett.42.673}. This framework established the foundational result that Anderson localization is especially significant in dimensions $\leq 2$ \cite{scaling_PhysRevLett.42.673}.

The same spirit is further generalized to the transfer matrix method (TMM) \cite{TMM_MacKinnon_1981,TMM_MacKinnon_1983,FSS_Pichard_1981}.
TMM is based on the fact that all one-dimensional (1D) systems are localized in the presence of disorder \cite{scaling_PhysRevLett.42.673}. In TMM, the Schrödinger equation for a quasi-1D system of width $M$ and length $L$ is reformulated as a recursive propagation along the longitudinal direction, as shown in Fig.\ref{fig:schematic_diagram}(c). The asymptotic growth of the total transfer matrix over length $L$ defines the finite-width localization length $\xi_M$, determined from the smallest positive Lyapunov exponent \cite{TMM_MacKinnon_1981,TMM_MacKinnon_1983,FSS_Pichard_1981}. Additional details on the TMM can be found in Appendix \ref{supp:TMM_details}. Finite-size scaling is then carried out using the dimensionless ratio $\Lambda_M = \xi_M / M$ \cite{TMM_MacKinnon_1981,TMM_MacKinnon_1983,FSS_Pichard_1981,TMM_MacKinnon_1994}, analyzed as a function of disorder strength $W$. As illustrated in Fig.\ref{fig:schematic_diagram}(d–e), $\xi_M$ generally decreases with increasing $W$. However, the scaling of $\Lambda_M$ with $M$ reveals two distinct behaviors: if $\Lambda_M$ increases with $M$, the system flows to an extended (metallic) phase in the thermodynamic limit (see Fig.\ref{fig:schematic_diagram}(d)), whereas a decreasing $\Lambda_M$ with $M$ signals a localized (insulating) phase (see Fig.\ref{fig:schematic_diagram}(e)). We will use this idea repeatedly in our following discussion.
In this paper, for clarity we denote the localization length of quasi-1D systems as $\xi_M$, with $M$ the system width, and omit the subscript for purely 1D systems, writing it simply a $\xi$.

Besides TMM, a variety of other important approaches have been developed to investigate Anderson localization. These include the coherent potential approximation (CPA) \cite{p_soven_67,b_velicky_68}, typical medium theory (TMT) \cite{TMT_Dobrosavljevic_2003}, and its cluster extension, the typical medium dynamical cluster approximation (TMDCA) \cite{tmdca_review}, as well as the localization criterion on the sensitivity to boundary conditions \cite{Boundary_Edwards_1972}, perturbation theory \cite{altshuler1985electron,Lee_RevModPhys.57.287}, the nonlinear $\sigma$ model \cite{nonlinear_efetov1980interaction,nonlinear_wegner,efetov1999supersymmetry,ALT_Evers_2008}, and the self-consistent theory of localization \cite{Self-consistent_PhysRevLett.45.842}. Numerical and statistical approaches such as exact diagonalization (ED) \cite{Schenk_2006}, the inverse participation ratio (IPR) \cite{IPR_Wegner_1980,IPR_Hikami_1986,IPR_Alexander_2000,ALT_Evers_2008}, energy-level statistics \cite{Hofstetter_1993,Shklovskii_1993,IPR_Alexander_2000,SUNTAJS_2021}, multifractal analysis \cite{JANSSEN19981,IPR_Mirlin_2000,ALT_Rodriguez_2011}, and entanglement entropy \cite{EE_refael2004entanglement,EE_laflorencie2005scaling,EE_bardarson2012unbounded,EE_berkovits2012entanglement,EE_bauer2013area,EE_Zhao_2013,EE_pouranvari2014area} have also provided valuable insights, while Wannier function analysis \cite{Wannier_Resta_1999,Wannier_Resta_2011,Wannier_Marzari_2012} offers an alternative perspective on localization phenomena. All these methods lead to a comprehensive understanding of Anderson localization \cite{AL50}.

On the other hand, the density matrix is a powerful formalism in quantum mechanics and quantum statistical mechanics \cite{landau2013quantum}.
Mathematically, the density matrix encodes all measurable information about a system, enabling the calculation of probabilities and correlations. In many-body physics and condensed matter theory, the density matrix plays a central role in understanding transport properties, entanglement structure, and quantum statistical mechanics. Its versatility makes it an indispensable tool across fields ranging from quantum information to strongly correlated electron systems \cite{DMRG_PhysRevLett.69.2863,DMRG_RevModPhys.77.259}.
It is natural to ask whether the density matrix can also reveal information about localization. As discussed above, the TMM extracts localization through the localization length $\xi_M$. Thus, the central question we aim to address in this work is how the localization length can be directly obtained from the density matrix.

The paper is organized as follows. In Sec.~\ref{sec:non-inteacing}, we present the one-particle density matrix approach for non-interacting disorder systems
 via the modular density matrix. Then, we benchmark it against the TMM for both the 3D Anderson model (Sec.~\ref{sec:3d_anderson}) and the 2D spin-orbit coupled Anderson model (Sec.~\ref{sec:2d_soc_anderson}), finding exact agreement in both localization lengths and critical disorder values. Sec.~\ref{sec:multi_orbital_anderson} extends the method to a multi-orbital setting and benchmarks it against TMDCA. In Sec.~\ref{sec:interacting}, we generalize the modular density matrix to interacting systems through the many-body subtraction density matrix, verifying its consistency in the 1D spinless interacting fermion model (Sec.~\ref{sec:spinless_interacting}). More importantly, in Sec.~\ref{sec:anderson_hubbard_interacting} we uncover a correlated metallic phase in the 2D Anderson-Hubbard model at $n=4/15$. The summary and outlook are given in Sec.~\ref{sec:summary_and_outlook}.

\section{Non-interacting System}
\label{sec:non-inteacing}

In this section, we focus on the Anderson localization problem in non-interacting systems.
To illustrate our approach more clearly, we begin with simple, intuitive examples by considering a one-dimensional chain. An extended state with momentum $k$ can be written as $|\psi_k\rangle = \sum_{i} e^{ikx_i} c_{i}^\dagger | \text{vac} \rangle $
while a localized state may be represented as $|\psi_l\rangle = \sum_{i} e^{-x_i/\xi} c_{i}^\dagger | \text{vac} \rangle $, where the normalization factors are omitted for simplicity. From these states, the one-particle density matrix can be evaluated through $\rho(x) = \langle \psi | c_0^\dagger c_x | \psi \rangle$ between site $0$ and site at $x$. Then, we have 
\begin{eqnarray}
    \rho(x) \propto
\begin{cases}
     e^{ikx}, & \text{ extended}\\
     e^{-x/\xi},       & \text{ localized}
\end{cases}
\end{eqnarray}
Thus, for localized wavefunctions, the density matrix naturally contains the information about the localization length $\xi$.

Building on this observation and the scaling philosophy of the TMM, we propose a generalized modular density matrix (MDM) method to extract the localization length in quasi-1D systems. The overall procedure is illustrated in Fig.\ref{fig:flow_diagram}.
As an example, we consider the quasi-1D single-orbital Anderson model described by the spinless fermion Hamiltonian
\bea
\label{eq:anderson_model}
H=-t\sum_{\langle(i,\alpha),(j,\beta)\rangle}c_{i,\alpha}^\dagger c_{j,\beta} + \sum_{(i,\alpha)} \epsilon_{i,\alpha} n_{i,\alpha}
\eea
where $t$ denotes the nearest-neighbor hopping amplitude and $\epsilon_{i,\alpha}$ is the onsite random potential, uniformly distributed within $[-W, W]$. $i$ and $j$ index slices along the longitudinal direction, and $\alpha,\beta$ denote internal site index within a slice, as illustrated in Fig.\ref{fig:schematic_diagram}(c).

We first describe the computational workflow in the simplest case of a purely 1D chain, where each slice contains a single site (i.e., $\alpha=\beta=1$). We begin by generating $N_s$ disorder realizations of such 1D chains with length $L$  (Fig.~\ref{fig:flow_diagram}(a)). Each sample is diagonalized to obtain eigenenergies $E_n$ and eigenstates $|\psi_{n}\rangle$ ($n=1,2,\dots,L$). For each sample, we select the single-particle eigenstate $|\psi_{s,E_F}\rangle$ ($s$ labeling the index of a disorder realization) whose energy is closest to the Fermi level $E_F$, as illustrated in Fig.~\ref{fig:flow_diagram}(b). Since only this eigenstate is needed, Lanczos-based sparse diagonalization \cite{Lanczos_lanczos1950iteration,Lanczos_saad2011numerical} can be used to efficiently extract it.

For quasi-1D systems with internal degrees of freedom ($\alpha,\beta$) within each slice, it follows the same computational procedure to obtain $|\psi_{s,E_F}\rangle$ as illustrated above for the 1D case. 
From these states, we define the general MDM in the quasi-1D case as

\bea
\label{eq:MDM_definition}
\rho^m_{\alpha\beta}(x)\coloneqq\frac{1}{\mathcal{N}}\sum_{s=1}^{N_s}\sum_i |\bra{\psi_{s,E_F}}c_{i,\alpha}^\dagger c_{i+x,\beta}\ket{\psi_{s,E_F}}|
\eea
where $\mathcal{N}$ is a normalization factor and $\rho^m_{\alpha\beta}(x)$ denotes the $(\alpha,\beta)$-th matrix element of $\rho^m(x)$. The dimension of $\rho^m(x)$ corresponds to the total number of sites per slice. To extract the slowest decaying channel of $\rho^m(x)$, we further define
\bea
\label{eq:symmetrization}
\gamma(x) \coloneqq \lambda_{\max}\;[\rho^{m\dagger}(x)\,\rho^{m}(x)]
\eea
where $\lambda_{\max}[\cdot]$ denotes the largest eigenvalue of the Hermitian matrix $\rho^{m\dagger}(x)\rho^{m}(x)$. In the simplest case of a 1D single-orbital chain, the MDM reduces to a scalar with $\gamma(x) = |\rho^m(x)|^2$. The symmetrization guarantees that $\gamma(x)$ is real and positive, eliminating ambiguities from complex phases.

Physically, $\gamma(x)$ represents the slowest decaying mode of $\rho^m(x)$, directly analogous to the smallest positive Lyapunov exponent in TMM \cite{TMM_MacKinnon_1981,TMM_MacKinnon_1983,FSS_Pichard_1981}. After averaging over slices and disorder realizations, $\gamma(x)$ exhibits the expected exponential decay, $\gamma(x) = A\, e^{-x/\xi} ,$ from which the localization length $\xi$ can be directly extracted. An example for a 1D non-interacting chain with $W = 2t$ is shown in Fig.\ref{fig:flow_diagram}(c), where $\xi$ is determined to be $6.08$. The term “modular” in MDM highlights that the modulus of the one-particle density matrix is used as the basic building block, while averaging over sites and disorder samples guarantees statistical stability and robustness against randomness. Taking the modulus is crucial, as the raw amplitudes contain sign and phase oscillations; only after this operation does the resulting quantity reveal a clear exponential decay from the wavefunction envelopes. The connections between the MDM approach and the TMM framework are discussed in detail in Appendix \ref{supp:method_comparison}.

\subsection{MIT in 3D Anderson model}
\label{sec:3d_anderson}
\begin{figure*}[htbp]
    \centering
    \includegraphics[width=1.0\linewidth]{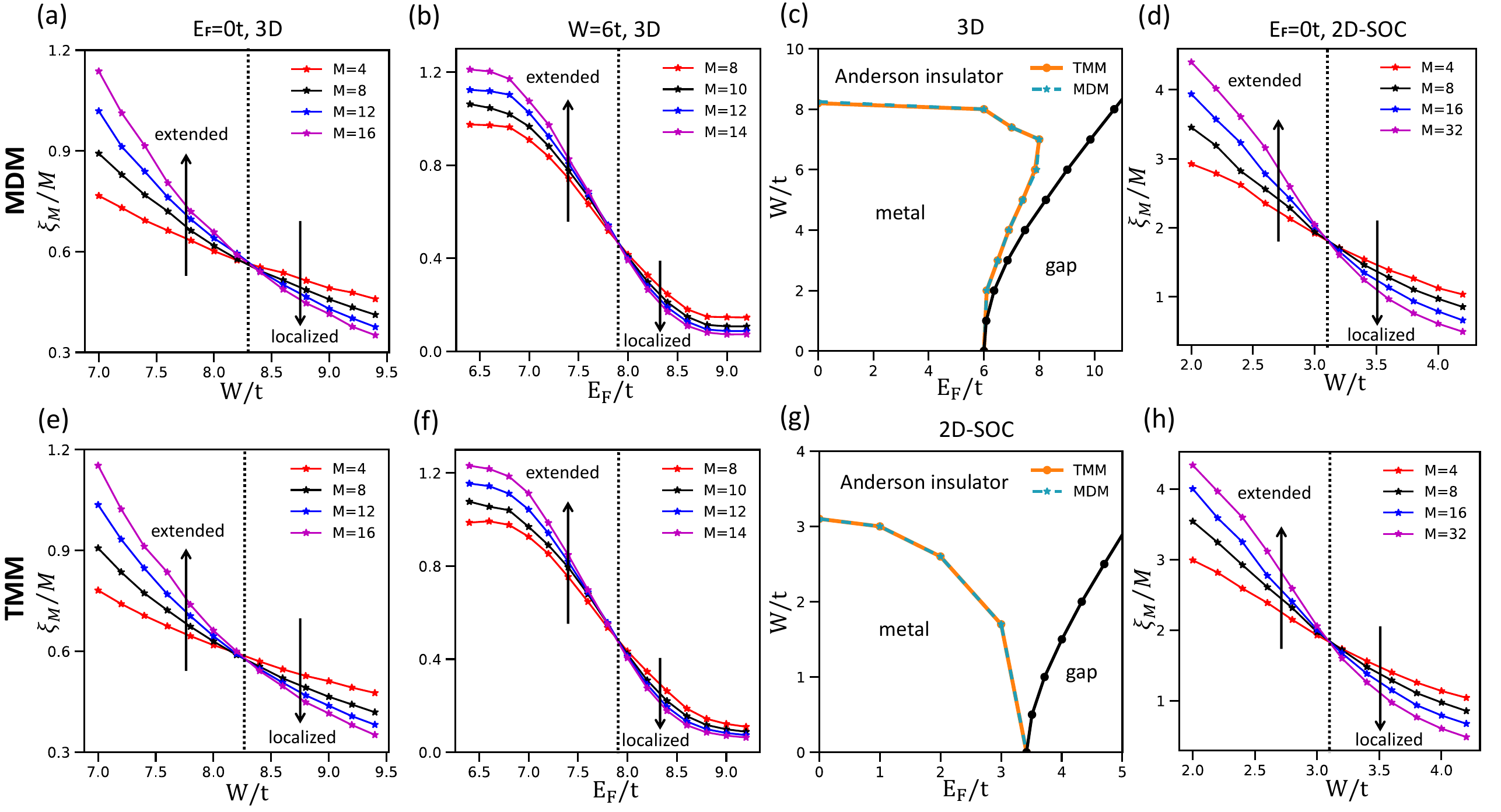}
    \caption{ Comparison of localization length scaling between MDM and TMM. (a)-(b) are finite-size scaling of $\xi_M/M$ obtained from MDM for the 3D Anderson model at $E_F=0$ as a function of $W$ (a) and at a fixed disorder strength $W=6t$ as a function of $E_F$ (b). (e)-(f) are the corresponding results from the TMM. Black dashed lines mark the metal-insulator transition points. (c), (g) are the phase diagrams of the 3D Anderson model and the spinful 2D Anderson model with SOC from both methods, respectively. (d), (h) are finite-size scaling of $\xi_M/M$ for the spinful 2D Anderson model with SOC at $E_F=0$ from MDM and TMM, respectively. Each $\xi_M/M$ data point from the MDM is obtained by averaging over 3000 disorder realizations of size $M^{d-1}\times L$ with $L=300$ for 3D Anderson model and $L=500$ for 2D Anderson model with SOC ($d$ is the dimension of the considered model), while in TMM each data point is obtained from quasi-1D systems of length $L=3\times10^6$ by averaging over three disorder realizations. In (c) and (g), the black dot lines indicate the energy spectral boundaries, determined as stable values of the largest eigenenergy averaged over 50 disorder realizations of size $L^d$ with $L=100$. }
    \label{fig:TM_DM_compare}
\end{figure*}

To benchmark our method, we start to apply the MDM approach to the standard 3D Anderson model. This model is defined on a cubic lattice with nearest-neighbor hopping and onsite random potential, taking the same quasi-1D Hamiltonian form as in Eq.~\ref{eq:anderson_model}. It is well established that the system undergoes a metal–insulator transition (MIT) at a critical disorder strength $W_c \approx 8t \sim 8.5t$ for $E_F = 0$ \cite{TMM_MacKinnon_1994,FSS_Slevin_1999,ALT_Rodriguez_2011,ALT_Slevin_2014}, and also exhibits an energy-dependent mobility edge \cite{mobility_edge_Mott01011967,mobility_edge_Mott_1987,Mobility_Bulka_1985,Mobility_Bulka_1987}.

We first focus on the case $E_F=0$ by tuning the disorder strength $W$. As shown in Fig.~\ref{fig:TM_DM_compare}(a), the localization length can be extracted from $\gamma(x)$ and analyzed using finite-size scaling within the MDM framework. Two distinct regimes are clearly observed: for $W < W_c \approx 8.3t$, the ratio $\xi_M/M$ increases with system width $M$, signaling extended states; whereas for $W > W_c \approx 8.3t$, $\xi_M/M$ decreases with $M$, indicating localized states. This distinct behavior is the hallmark of MIT. For comparison, we also perform the TMM calculation, shown in Fig.~\ref{fig:TM_DM_compare}(e). Remarkably, the value of localization lengths $\xi_M$ and the critical disorder strength $W_c$ obtained from MDM are in excellent agreement with those from TMM.

Another characteristic feature of the Anderson model is the presence of a mobility edge, which separates extended and localized states depending on energy \cite{mobility_edge_Mott_1987,mobility_edge_Mott01011967}. To capture this, we vary $E_F$ at a fixed disorder strength $W$. The results from MDM for the 3D Anderson model at $W=6t$ are shown in Fig.~\ref{fig:TM_DM_compare}(b). Again, two distinct scaling regimes appear: states near the band center (small $E_F$) remain extended, while those near the band edges (large $E_F$) become localized. This behavior is also consistent with the TMM results presented in Fig.~\ref{fig:TM_DM_compare}(f).

We can now determine the phase diagram of the 3D Anderson model \cite{TMM_MacKinnon_1994,FSS_Slevin_1999,ALT_Slevin_2014,ALT_Rodriguez_2011,Mobility_Bulka_1985,Mobility_Bulka_1987}, as shown in Fig.~\ref{fig:TM_DM_compare}(c).
At $W=0$, the tight-binding band edge lies at $6t$, separating the metallic phase from the fully gapped region.
As disorder strength $W$ increases, the band edges broaden as expected  \cite{THOULESS197493}, forming the upper boundary indicated by the black dotted line in Fig.~\ref{fig:TM_DM_compare}(c). The black dots are  obtained from the stabilized values of the largest eigenenergy averaged over 50 disorder realizations on lattices of size $L^d$ with $L=100$.
Between the metallic region and the gapped region lies the Anderson insulating phase, which has a finite density of states at $E_F$ but remains insulating due to localization.
The phase boundary separating the metallic and Anderson insulating phases is obtained independently from both MDM and TMM, with the two methods showing nearly identical results.
Hence, our method offers a robust and precise characterization of the 3D Anderson model throughout the entire phase space. Additional numerical results and analysis of the 3D Anderson model can be found in Appendix \ref{supp:3D-anderson}.

\subsection{MIT in spinful 2D Anderson model with SOC} 
\label{sec:2d_soc_anderson}

According to the single-parameter $g(L)$ scaling theory, all states in a two-dimensional (2D) disordered system are localized  \cite{scaling_PhysRevLett.42.673}.
This conclusion, however, is contingent on the system belonging to the orthogonal symmetry class \cite{SOC_10.1143/PTP.63.707}. In the presence of spin-orbit coupling (SOC), spin-rotation symmetry is broken while time-reversal symmetry is preserved, which fundamentally alters the system's classification by moving it into the symplectic class \cite{SOC_10.1143/PTP.63.707,BERGMANN19841_soc,symmetry_PhysRevB.55.1142,ALT_Evers_2008,efetov1999supersymmetry}. The key physical manifestation of this symmetry change is weak anti-localization (WAL), where destructive quantum interference between time-reversed paths suppresses backscattering \cite{BERGMANN19841_soc}. Because this effect counteracts localization, a 2D system in the symplectic class can undergo an Anderson metal-insulator transition, a phenomenon forbidden in the standard orthogonal case.

Given the distinctive localization behavior induced by SOC, we revisit the Anderson metal–insulator transition in a 2D system with SOC using the MDM framework. In particular, we focus on the SU(2) model proposed in Ref.~\cite{SOC_Asada_2002}, which describes a spinful Anderson system on a square lattice with both random onsite potentials and random SOC. In the quasi-1D system of size $M\times L$, the Hamiltonian of the 2D SU(2) model can be written as
\bea
\label{eq:2D_SOC}
H_{SOC} = 
- \sum_{\langle (i,\alpha),(j,\beta)\rangle}\sum_{\sigma,\sigma'}
t\, R_{(i,\alpha);(j,\beta)}^{\sigma\sigma'}\, 
c^{\dagger}_{i,\alpha,\sigma} c_{j,\beta,\sigma'}+\sum_{(i,\alpha)}\sum_{\sigma} 
\epsilon_{i,\alpha}\, n_{i,\alpha,\sigma}
\eea

Here, $i$, $j$ are the slice indices and the $\alpha$,$\beta$ are the site indices within a slice following the same convention as that in Eq.\ref{eq:anderson_model}. The additional ingredient is the spin degree of freedom labeled by $\sigma$,$\sigma'\in \{\uparrow,\downarrow\}$. Accordingly, in the spinful case, the MDM in Eq.\ref{eq:MDM_definition} is extended to include the spin indices, such that its elements are written as $\rho^m_{\alpha\sigma,\beta\sigma'}(x)$, corresponding to $|\bra{\psi_{s,E_F}}c_{i,\alpha,\sigma}^\dagger c_{i+x,\beta,\sigma'}\ket{\psi_{s,E_F}}|$ averaged over disorder realizations ($s$) and slices ($i$).
The onsite disorder $\epsilon_{i,\alpha}$ is uniformly distributed in $[-W,W]$ and the random $2\times2$ unitary matrices $R_{(i,\alpha);(j,\beta)}^{\sigma\sigma'}$ represent the SOC terms on  nearest-neighbor bonds  (see Ref.~\cite{SOC_Asada_2002} and Appendix \ref{supp:2D-anderson-SOC} for the definition). This model preserves time-reversal symmetry but breaks spin-rotation symmetry due to SOC, placing the model in the symplectic universality class.

Fig.\ref{fig:TM_DM_compare}(g) shows the phase diagram of the SU(2) model obtained from both MDM and TMM.
There are also three different phases: metal, Anderson insulator, and gap phases.
Same as the results for the 3D Anderson model presented in the previous subsection, the metal–insulator phase boundary determined by MDM coincides with that from TMM. 
We also explore the scaling of $\xi_M/M$ at $E_F=0$, displayed in Fig.\ref{fig:TM_DM_compare}(d). For disorder strength below $W_c \approx 3.1t$, the system remains delocalized, while for $W > W_c \approx 3.1t$ it becomes localized. The corresponding TMM results, shown in Fig.\ref{fig:TM_DM_compare}(h), exhibit the same trend.
In contrast, the 2D orthogonal case does not exhibit a metal–insulator transition, as discussed in the Appendix \ref{supp:2D-anderson-SOC}.  
Together, these findings confirm that the MDM framework provides an accurate and robust description even in systems with spin degree of freedom and complex hopping terms, thereby establishing its applicability to Anderson transitions across different symmetry classes. Additional numerical results and analysis of the 2D Anderson model with SOC can be found in Appendix \ref{supp:2D-anderson-SOC}.

\subsection{MIT in Multiorbital systems}
\label{sec:multi_orbital_anderson}
\begin{figure}[htbp]
    \centering
    \includegraphics[width=1.01\linewidth]{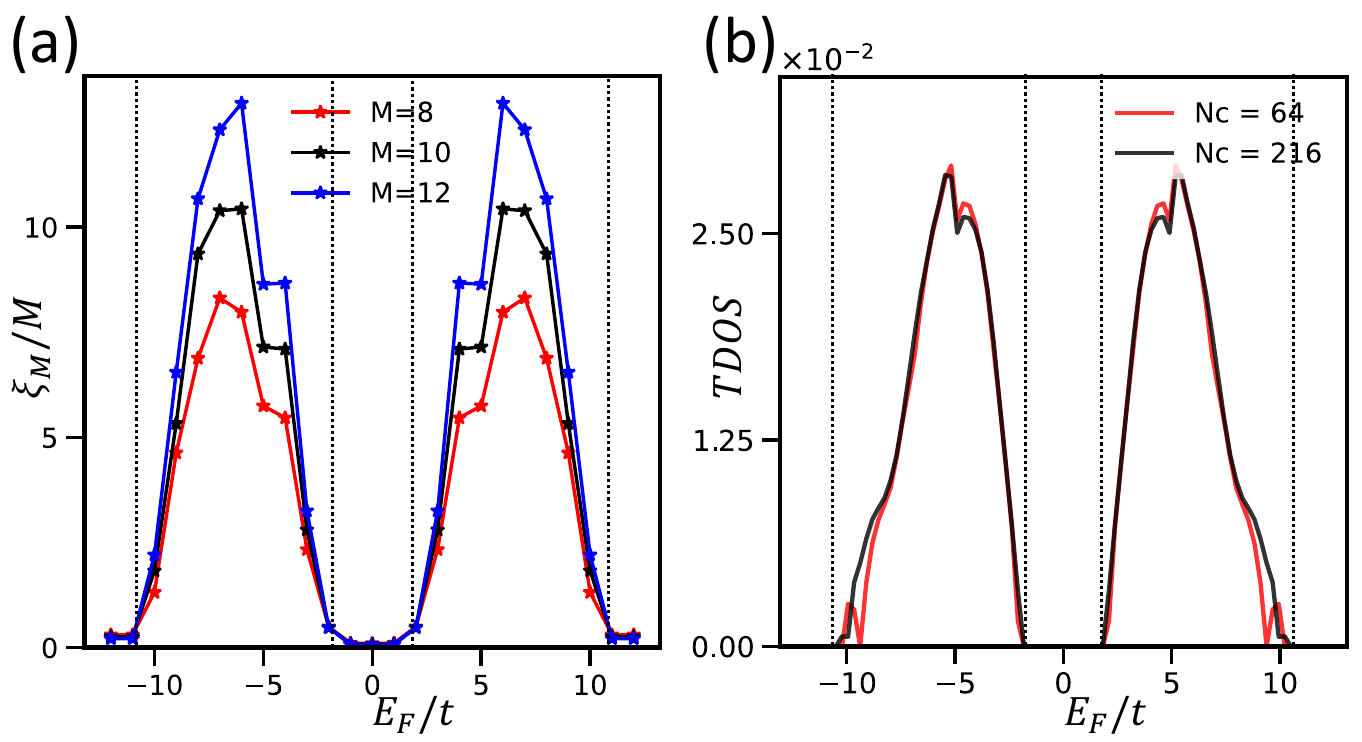}
    \caption{ Comparison of results from MDM and TMDCA for the two-orbital 3D Anderson model at a fix disorder strength of $W=5t$. (a) Finite-size scaling of $\xi_M/M$ obtained from MDM, with system size $M^{d-1}\times L\times2$ ($L=300$) averaged over 3000 disorder realizations. (b) TDOS obtained from TMDCA with cluster sizes $N_c=216$ (black) and $N_c=64$ (red). Model parameters are $t^{11}=t^{22}=t$, $t^{12}=t^{21}=0.3t$ and $W_A=-W_B=5t$. Vertical dashed lines in both panels mark the phase boundary between localized and extended states.}
    \label{fig:multi_orbital}
\end{figure}

Beyond the single-orbital limit discussed above, realistic materials inevitably involve multi-orbital physics. With several orbitals near the Fermi level, inter-orbital hybridization and nonlocal disorder effects naturally arise. 
It is therefore essential to establish that our method can accurately capture the localization transition in such multi-orbital settings. Demonstrating this capability not only validates the robustness of our approach but also opens the path toward its integration with first-principles electronic structure methods, enabling realistic investigations of disorder-driven localization in complex materials.
In this subsection, we benchmark MDM against TMDCA in multiorbital models. 
TMDCA \cite{tmdca_review} is a cluster extension of TMT \cite{TMT_Dobrosavljevic_2003}, which generalizes the widely used CPA \cite{p_soven_67,b_velicky_68} by replacing the arithmetic average of the local density of states (LDOS) in CPA with its geometric average. The resulting average density of states approximates the typical value of the LDOS, denoted as TDOS, which can serves as an order parameter for the Anderson localization transition \cite{JANSSEN19981,Byczuk_2010,hanna_2014, TMDCA_Zhang_2015,ekuma_2015,zhangyi_2016,zhang_2018}.

Specifically, we consider a two-orbital Anderson model studied in Ref.~\cite{TMDCA_Zhang_2015}, defined on a cubic lattice with intra-orbital binary disorder. The Hamiltonian of this two-orbital model ($H_{TO}$) in the quasi-1D system of length $L$ and width $M$ can be written as 
\bea
\label{eq:multi_orbital}
H_{TO} = - \sum_{\mu,\nu=1}^2 \sum_{\langle (i,\alpha),(j,\beta)\rangle} t^{\mu\nu}
c^{\dagger}_{i,\alpha,\mu} c_{j,\beta,\nu}
+ \sum_{\mu=1}^{2}\sum_{(i,\alpha)}
\epsilon^{\mu}_{i,\alpha}\, n_{i,\alpha,\mu}
\eea
Here, $\mu$,$\nu\in\{1,2\}$ denote the orbital degree of freedom, the hopping amplitudes $t^{\mu\nu}$ include both intra-orbital ($\mu=\nu$) and inter-orbital ($\mu\neq\nu$) terms. The onsite disorder $\epsilon^{\mu}_{i,\alpha}$ acts only within each orbital channel and follows a binary distribution $P(\epsilon^{\mu}_{i,\alpha})=\frac{1}{2}\delta(\epsilon^{\mu}_{i,\alpha}-W_A)+\frac{1}{2}\delta(\epsilon^{\mu}_{i,\alpha}-W_B)$, so that each orbital independently takes the value of random potential as $W_A$ or $W_B$ with equal probability. For the MDM, similar to the case of the spinful Anderson model, the matrix elements of $\rho^m(x)$ for the multi-orbital Anderson model need to carry the additional orbital degree of freedom as $\rho^m_{\alpha\mu,\beta\nu}(x)$, ensuring that both spatial and orbital degrees of freedom are retained.

Compared to the single-orbital Anderson model with a uniform box distribution, this two-orbital model exhibits a distinct phase diagram \cite{TMDCA_Zhang_2015}. The binary distribution of disorder sharpens the separation between extended and localized states, while inter-orbital hopping broadens the effective bandwidth and shifts the mobility edge to higher disorder strengths. As the disorder strength increases, both states near the band center and the band edges begin to localize.

To benchmark our approach, we focus on the scaling of the localization length extracted from MDM at a fixed disorder strength of $W_A=-W_B=5t$,  while scanning the Fermi energy $E_F$. The scaling results in Fig.~\ref{fig:multi_orbital}(a) show that at this disorder strength, the two-orbital model hosts two particle-hole symmetric metallic regions bounded by finite critical energies, separated by a localized phase around the band center.  Thus, there are two pairs of phase boundaries, ($\pm E_{Fc1}$, $\pm E_{Fc2}$), separating metallic and insulating states. This behavior is reflected in the scaling of $\xi_M/M$, which increases with width $M$ in the metallic phase and decreases in the localized phase. The corresponding TMDCA results in Fig.\ref{fig:multi_orbital}(b) shows finite (vanishing) TDOS in the same metallic (localized) regions as that in Fig.\ref{fig:multi_orbital}(a), which serves as the order parameter for the Anderson localization. The agreement between the two methods is excellent, with nearly identical phase boundaries at $E_{Fc1}\approx \pm1.8t$ and $E_{Fc2}\approx \pm10.5t$, as indicated by the vertical dashed lines in Fig.~\ref{fig:multi_orbital}. This demonstrates that MDM faithfully reproduces the unique mobility-edge structure of this multi-orbital system. Importantly, it also suggests that the MDM framework can be extended beyond toy models to realistic multi-orbital materials, where orbital complexity plays a crucial role.

Generalizations of localization theory to realistic materials have already been successfully implemented in frameworks such as the TMT \cite{TMT_Dobrosavljevic_2003} and TMDCA \cite{ekuma_2015}, both of which can be seamlessly combined with dynamical mean-field theory (DMFT) for strongly correlated systems \cite{TMT_DMFT_PhysRevLett.94.056404,Nguyen_2022}.
Similarly, a natural extension of our approach points toward density matrix embedding theory (DMET) \cite{DMET_PhysRevLett.109.186404,dmet_guide}. Unlike DMFT, which is formulated around the local Green’s function, the central object in DMET is the frequency-independent local density matrix. By enforcing a self-consistent match between an impurity cluster and its environment bath, DMET achieves an accurate description of strongly correlated systems directly through the density matrix. This structural similarity suggests that our framework could be seamlessly embedded into DMET, opening a path toward studying localization in realistic multiorbital and strongly correlated materials. We leave this promising direction for future work.

\section{Interaction System}
\label{sec:interacting}
\begin{figure*}[htbp]
    \centering
    \includegraphics[width=1.00\linewidth]{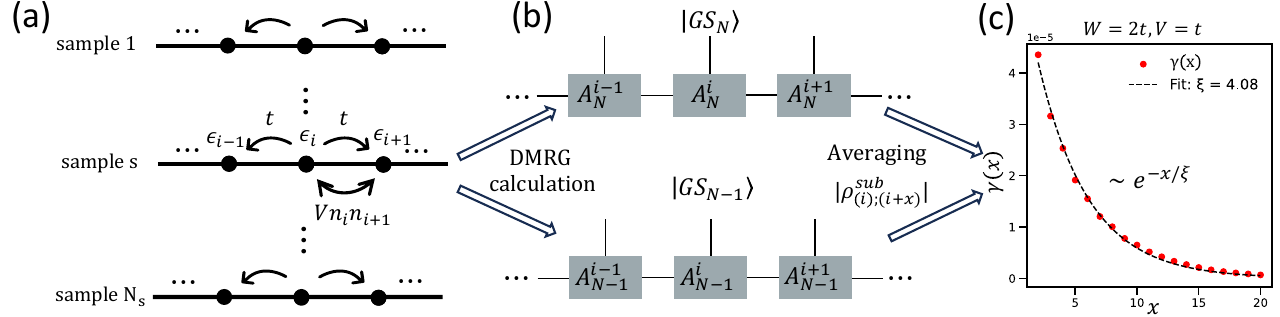}
    \caption{Computational workflow for extracting the localization length using the many-body extension of MDM approach in spinless 1D interacting systems. (a) schematically shows different disorder samples with hopping energy $t$, onsite random potential $\epsilon_i$ and nearest-neighbor electron-electron interaction $V$. %After performing diagonalization on 
    For each disorder sample, $\ket{GS_N}$ and $\ket{GS_{N-1}}$ are obtained by performing DMRG calculation, as illustrated in (b). Then the SDM $\rho^{sub}_{(i);(i+x)}$ can be obtained from the two states according to Eq.\ref{eq:SDM_definition}. $A_N^i$ ($A_{N-1}^i$) represents the matrix product state of $\ket{GS_N}$ ($\ket{GS_{N-1}}$) at site $i$. (c) shows $\gamma(x)$ as a function of $x$ for the systems with $W=2t$ and $V=t$, which exhibits clear exponential decay with a localization length $\xi=4.08$. (c) is calculated by averaging 500 disorder realizations of length $L=120$ under open boundary condition at half-filling. In the site averaging for each sample, the 20 sites nearest to each boundary are excluded to minimize open-boundary effects.}
    \label{fig:flow_diagram_Manybody}
\end{figure*}

The theory of Anderson localization in non-interacting systems has been extensively developed over the past sixty years \cite{THOULESS197493,Lee_RevModPhys.57.287, AL50}, culminating in a well-established scaling framework. Once electron–electron interactions are included, however, the problem becomes substantially more complex, since interactions can both compete with and enhance disorder effects. Early theoretical progress was made using diagrammatic perturbation theory, which revealed that quantum interference, combined with interactions, leads to nontrivial corrections to conductivity and thermodynamic quantities \cite{altshuler1985electron,ALTSHULER1979115,Altshuler1979ContributionTT,Altshuler_PhysRevLett.44.1288,Finkelshtein_1983,Castellani_PhysRevB.30.527}. These approaches laid the groundwork for understanding interaction-induced dephasing, zero-bias anomalies, and Altshuler–Aronov-type corrections \cite{Lee_RevModPhys.57.287}.
More recently, the study of interacting disordered systems has been revitalized by the study of many-body localization (MBL) \cite{MBL_PhysRevB.21.2366,MBL_BASKO20061126,MBL_RevModPhys.91.021001}, which generalizes Anderson localization to finite energy densities. In this regime, interactions fail to restore ergodicity, leading to localized many-body eigenstates characterized by emergent local integrals of motion. One-particle density matrix has been used to study MBL in previous works \cite{MBL_DM_1,MBL_DM_2,MBL_DM_3}, capturing unique information of many-body localized eigenstates.

Despite these advances, a comprehensive and unified theoretical framework for the interplay between disorder and interactions remains elusive. A central question concerns \textit{whether low-dimensional electronic systems can sustain metallic phases}. The scaling theory of localization predicts the absence of true metallic behavior in 2D \cite{scaling_PhysRevLett.42.673}. However, this long-standing conclusion has been challenged by experiments on 2D electron systems, which provide compelling evidence that an MIT can indeed occur in 2D \cite{MIT_2D_Kravchenko_2004,MIT_2D_PhysRevB.50.8039,MIT_2D_PhysRevB.51.7038,MIT_2D_PhysRevLett.77.4938,MIT_2D-np,MIT_2D_Rev_2000}.
Many attempts have been made to address this issue, including determinant quantum Monte Carlo \cite{DQMC_PhysRevLett.83.4610,DQMC_PhysRevLett.87.146401,DQMC_PhysRevLett.93.126401,DQMC_PhysRevB.84.035121}, 
zero-temperature Green function quantum Monte Carlo \cite{QMC_PhysRevLett.101.226803}, projector quantum Monte Carlo \cite{PGQMC_PhysRevB.67.205112}, non-linear sigma model \cite{Finkel'stein-science} and Hatree-Fock calculation \cite{HF_PhysRevLett.81.4212,HF_PhysRevLett.93.126401}.
Here, we aim to approach this problem from the perspective of the density matrix.

However, electron–electron correlations make the diagonalization scheme used in non-interacting systems infeasible, making it impossible to directly obtain the single-particle wavefunction $\ket{\psi_{E_F}}$ in Eq.\ref{eq:MDM_definition}. Instead, one must work with many-body wavefunctions. To address this issue, we extend our approach to the interacting case by formulating a many-body version of MDM. Specifically, we introduce a one-particle subtraction density matrix (SDM) $\rho_{(i,\alpha);(j,\beta)}^{sub}$, defined from the difference between two many-body ground states as
\bea
\label{eq:SDM_definition}
    \rho_{(i,\alpha);(j,\beta)}^{sub}=\langle GS_{N}|c_{i,\alpha}^\dagger c_{j,\beta}|GS_{N}\rangle-\langle GS_{N-1}|c_{i,\alpha}^\dagger c_{j,\beta}|GS_{N-1}\rangle 
\eea
Where $|GS_{N}\rangle$, $|GS_{N-1}\rangle$ denote the ground state wavefunctions with fixed particle numbers $N$ and $N-1$, respectively. The notations $i$,$j$ label the slice indices along the longitudinal direction, while $\alpha$,$\beta$ denote site indices within a slice, following the same convention as in Eq.\ref{eq:MDM_definition}. 
The diagonal terms of SDM with $(i,\alpha)=(j,\beta)$
have been initially used in quantum chemistry \cite{SDM_Fuku} and quasi-periodic 1D systems \cite{SDM_anderson} to capture reactive/ionization hot spots and critical localization respectively.
In our work, we present the use of the off-diagonal terms of the SDM to extract the localization length in interacting many-body systems.

To clarify above definition, we first examine $\rho_{(i,\alpha);(j,\beta)}^{sub}$ in the non-interacting limit. In this case, the many-body ground states $|GS_{N}\rangle$ and $|GS_{N-1}\rangle$ are product states of orthogonal single-particle eigenstates $\ket{\psi_{n}}$ with eigenenergy $E_n$ by
\bea
|GS_{N}\rangle = \prod_{n=1}^N |\psi_{n}\rangle
= |\psi_{E_F}\rangle \otimes \prod_{n=1}^{N-1}|\psi_{n}\rangle
= |\psi_{E_F}\rangle \otimes |GS_{N-1}\rangle
\eea
%Here, $E_F$ corresponds to the energy of the highest occupied single-particle eigenstate in $|GS_{N}\rangle$.
Using the orthogonality of single-particle eigenstates, one can show that $\rho_{(i,\alpha);(j,\beta)}^{sub}$ reduces to $\bra{\psi_{E_F}}c_{i,\alpha}^\dagger c_{j,\beta}\ket{\psi_{E_F}}$, which is exactly the basic building block of MDM we used in the non-interacting limit (see Eq.\ref{eq:MDM_definition}). 

We can further go beyond this product state assumption. Suppose we have two normalized ground state wavefunctions $|GS_{N}\rangle$ and $|GS_{N-1}\rangle$ connected by a fermionic operator $\hat{\psi}$ as $|GS_{N}\rangle=\hat{\psi}^\dagger |GS_{N-1}\rangle$. Then, we can prove that $\rho_{(i,\alpha);(j,\beta)}^{sub}$ just leads to the information of $\hat{\psi}$.
%More generally, we now consider the case beyond simple product states. 
More precisely, we can assume that $\hat{\psi}$ has the form $\sum_{i,\alpha}a_{i,\alpha} c_{i,\alpha}$, where the normalization requires $\sum_{i,\alpha} | a_{i,\alpha}|^2=1$. Using the anticommutation relations $\{ \hat{\psi},c_{i,\alpha}^\dagger\}=a_{i,\alpha}$ and $\{ \hat{\psi}^\dagger,c_{j,\beta}\}=a_{j,\beta}^*$, the SDM defined in Eq.\ref{eq:SDM_definition} can be rewritten as 
\bea
\label{eq:SDM_derivation}
\rho_{(i,\alpha);(j,\beta)}^{sub}=&\langle  \hat{\psi} c_{i,\alpha}^\dagger c_{j,\beta} \hat{\psi}^\dagger \rangle_{N-1}-\langle c_{i,\alpha}^\dagger c_{j,\beta} \rangle_{N-1} \\
=&\langle (a_{i,\alpha} -c_{i,\alpha}^\dagger \hat{\psi})( a_{j,\beta}^*-\hat{\psi}^\dagger c_{j,\beta})  \rangle_{N-1}-\langle c_{i,\alpha}^\dagger c_{j,\beta} \rangle_{N-1}\\
=&a_{i,\alpha} a_{j,\beta}^*-a_{i,\alpha} \langle  \hat{\psi}^\dagger c_{j,\beta} \rangle_{N-1} - a_{j,\beta}^* \langle  c_{i,\alpha}^\dagger\hat{\psi}  \rangle_{N-1}+\\
&\langle  c_{i,\alpha}^\dagger\hat{\psi}\hat{\psi}^\dagger c_{j,\beta} \rangle_{N-1}-\langle  c_{i,\alpha}^\dagger c_{j,\beta} \rangle_{N-1}\\
=& a_{i,\alpha}a_{j,\beta}^* 
\eea
Here, $\langle\cdots\rangle_{N-1}$ is a simplified notation of $\langle GS_{N-1}|\cdots|GS_{N-1}\rangle$. The final equality is obtained by using the anticommutation relation $\langle  c_{i,\alpha}^\dagger(\hat{\psi}\hat{\psi}^\dagger+\hat{\psi}^\dagger\hat{\psi}) \;c_{j,\beta} \rangle_{N-1}=\langle  c_{i,\alpha}^\dagger c_{j,\beta} \rangle_{N-1}$ and $\hat{\psi} |GS_{N-1}\rangle=\hat{\psi} \hat{\psi}|GS_{N}\rangle=0$. 
Thus, this many-body version of MDM extracts precisely the information connecting $| GS_{N}\rangle$ and $| GS_{N-1}\rangle$ as $\rho_{(i,\alpha);(j,\beta)}^{sub}=a_{i,\alpha}a_{j,\beta}^* =\bra{\psi}c_{i,\alpha}^\dagger c_{j,\beta} \ket{\psi}$, where $\ket{\psi}=\hat{\psi}^\dagger\ket{0}$.

Consequently, we can define the many-body version of MDM $\rho^m(x)$ in direct analogy with the non-interacting case by averaging the modulus of SDM over sites and disorder samples as
\bea
\label{eq:SDM_MDM}
\rho^m_{\alpha,\beta}(x)=\frac{1}{\mathcal{N}}\sum_{s=1}^{N_s}\sum_{i} | \rho_{s,(i,\alpha);(i+x,\beta)}^{sub} |
\eea
Here, the additional index $s$ denotes the disorder realizations. For the non-interacting case, this definition is strictly equivalent to Eq.~\ref{eq:MDM_definition}, and directly encodes the localization properties of the single-particle state $\ket{\psi_{E_F}}$. More generally, when the ground states with $N$ and $N-1$ particles, $\ket{GS_{N}}$ and $\ket{GS_{N-1}}$, are connected by a fermionic operator $\hat{\psi}$, the SDM captures the localization characteristics of $\hat{\psi}$.

It is important to note that in the presence of electron–electron interactions, the simple relation $\ket{GS_{N}}= \hat{\psi}^\dagger \ket{GS_{N-1}}$ does not hold in general. Nevertheless, for finite-size quasi-1D systems at finite disorder strength, the ground states are always localized except in special cases, and thus we expect to have this localized operator $\hat{\psi}$. While interactions may complicate the precise form of $\hat{\psi}$, the exponential decay of $\rho^m_{\alpha\beta}(x)$ is expected to persist, providing a robust measure of the localization length in strongly disordered interacting systems. More detailed considerations, including extensions of the SDM to spinful systems and higher-order corrections to $\hat{\psi}$, are discussed in the Appendix \ref{supp:spinful_SDM}.

\subsection{Computational workflow}
\label{sec:workflow_interacting}
With SDM, the computational workflow for extracting the localization length from the many-body version of MDM is summarized in Fig.\ref{fig:flow_diagram_Manybody}. Here, we use the spinless fermion model in 1D lattice with interaction and disorder as an example \cite{DMRG_1D_PhysRevB.72.024208,DMRG_1D_PhysRevLett.80.560}, which can be written as
\bea
\label{eq:attractive_model}
H_{V} = -t \sum_{\langle i,j\rangle} c_i^\dagger c_j
 +V \sum_i n_i n_{i+1}+\sum_i \epsilon_i n_i
\eea
Here, $t$ is the nearest-neighbor hopping amplitude, $V$ denotes the strength of nearest-neighbor interaction ($V>0$ for repulsion and $V<0$ for attraction), and the onsite random potentials $\epsilon_i$ are uniformly distributed within $[-W,W]$ as in non-interacting case (Eq.\ref{eq:anderson_model}).

We begin by generating $N_s$ independent disorder realizations of the model in Eq.\ref{eq:attractive_model}, as illustrated in Fig.\ref{fig:flow_diagram_Manybody}(a). For each disorder realization, we employ the density matrix renormalization group (DMRG) method \cite{DMRG_PhysRevLett.69.2863}, as implemented in ITensor package \cite{DMRG_Itensor_1,DMRG_Itensor_2}, to compute two ground states $\ket{GS_{N}}$ and $\ket{GS_{N-1}}$ under open boundary condition (OBC), as illustrated in Fig.\ref{fig:flow_diagram_Manybody}(b). The DMRG calculations for the spinless 1D model are performed with 330 steps of sweep and the number of kept states is increased gradually to 400, ensuring the convergence and the truncation error $\epsilon<10^{-10}$. 
From these two states, we can calculate the SDM $\rho_{(i,\alpha);(j,\beta)}^{sub}$ ($\alpha=\beta=1$ in spinless 1D chains) according to Eq.\ref{eq:SDM_definition}. We then construct the many-body version of MDM $\rho^m(x)$ by averaging the modulus of the SDM over sites and disorder samples as in Eq.\ref{eq:SDM_MDM}. After symmetrization, the largest eigenvalue of $\rho^m(x)\rho^{m\dagger}(x)$ defines $\gamma(x)$, following the same procedure in Eq.\ref{eq:symmetrization}. In the regime of sufficiently strong disorder, we find that $\gamma(x)$ still exhibits clear exponential decay even in the presence of electron-electron interactions. Fig.\ref{fig:flow_diagram_Manybody}(c) shows a representative example of disorder strength $W=2t$ and repulsive interaction $V=t$, where $\gamma(x)$ is well described by $e^{-x/\xi}$ with $\xi=4.08$. This result demonstrates that the many-body MDM precisely captures the localization information encoded in the operator $\hat{\psi}$ connecting $\ket{GS_{N}}$ and $\ket{GS_{N-1}}$. Further computational details of DMRG and benchmarks for individual samples with ED in the non-interacting case can be found in Appendix \ref{supp:benchmark}.

\begin{figure}[htbp]
    \centering
     \includegraphics[width=1.02\linewidth]{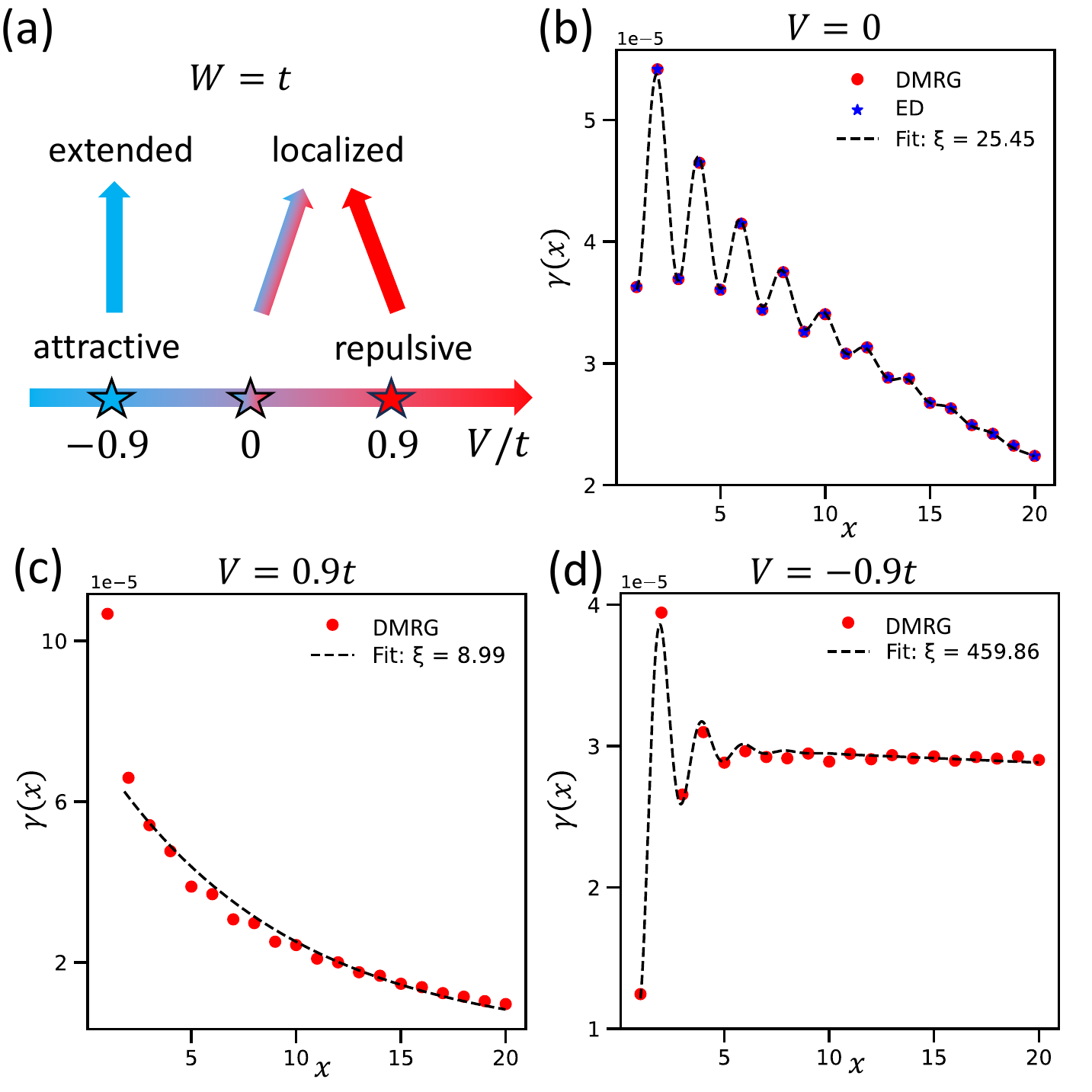}
    \caption{ Results for the 1D spinless interacting chain with finite onsite disorder strength $W=t$ at half-filling. (a) is a schematic phase diagram showing that for finite disorder strength ($W=t$), both the non-interacting case ($V=0$) and repulsive interactions ($V>0$) correspond to localized states, while at specific attractive interaction strengths ($V<0$) an extended state emerges.  (b)--(d) display $\gamma(x)$ and their fitting results (using Eq.\ref{eq:damp_oscillation}) for $V=0t$, $0.9t$, and $-0.9t$  The extracted parameters are $\xi=25.45,\;\xi_O=4.00$ for $V=0$; $\xi=8.99,\;\xi_O\to 0$ for $V=0.9t$; and $\xi=459.86,\;\xi_O=1.32$ for $V=-0.9t$. Each panel among (b)--(d) is obtained by averaging over 500 disorder realizations of length $L=120$ under OBC. In the site averaging for each sample, the 20 sites nearest to each boundary are excluded to minimize open-boundary effects. panel (b) is calculated using both ED and DMRG, while panel (c) and (d) are calculated by DMRG. 
}
    \label{fig:attractive_1D}
\end{figure}

\subsection{1D spinless interaction model}
\label{sec:spinless_interacting}
In one-dimensional correlated electron systems, it has long been proposed that the interplay between attractive interactions and Anderson localization can stabilize a delocalized phase \cite{giamarchi2003quantum,mattis1974new,Luther_PhysRevLett.32.992,Luther_PhysRevLett.33.589,W_Apel_1982,Apel_PhysRevB.26.7063,Giamarchi_PhysRevB.37.325,T_Giamarchi_1987}. For example, a delocalization transition occurs when the Luttinger parameter satisfies $K > 3/2$ in a spinless model within the bosonization framework \cite{giamarchi2003quantum}. Physically, this arises because attractive interactions enhance superconducting quantum fluctuations, which compete with disorder and can ultimately drive delocalization transition \cite{giamarchi2003quantum}. This prediction has been extensively tested through numerical studies using DMRG and related methods \cite{DMRG_1D_PhysRevLett.80.560,DMRG_1D_PhysRevB.72.024208,DMRG_1D_PhysRevB.75.064209,DMRG_1D_PhysRevB.87.205140,DMRG_1D_PhysRevLett.115.206401}.

To establish the validity of our method, we investigate how the interaction strength $V$ affects the localization properties in the spinless 1D interacting lattice model at half-filling using our many-body extension of the MDM approach. At relatively weak disorder ($W\sim t$), previous numerical calculations \cite{DMRG_1D_PhysRevLett.80.560,DMRG_1D_PhysRevB.72.024208,DMRG_1D_PhysRevB.75.064209,DMRG_1D_PhysRevB.87.205140,DMRG_1D_PhysRevLett.115.206401} have shown that while repulsive interactions ($V>0$) enhance localization compared to the non-interacting  case, attractive interactions ($V<0$) can induce nontrivial extended states in 1D, as illustrated in Fig.\ref{fig:attractive_1D}(a). To verify the accuracy of our DMRG calculation in the MDM approach, we first benchmark against the $V=0$ case, where the model reduces to the non-interacting Hamiltonian and the MDM can be obtained using both DMRG and non-interacting ED. 
Fig.\ref{fig:attractive_1D}(b) shows the results of $V=0$ and $W=1$, averaged over the same 500 disorder realizations in DMRG and ED, where DMRG and non-interacting ED yield identical $\gamma(x)$, confirming the reliability of our DMRG calculations. 

However, the weak $W$ complicates both the $\gamma(x)$ and the fitting procedure. It is worth noting that in the clean limit ($W=V=0$) of a 1D nearest-neighbor chain, $\gamma(x)$ can be computed exactly and follows an oscillatory form determined by the Fermi momentum $k_F$ as $\gamma(x)\sim(1+\cos(2k_Fx))$. As disorder increases, $\gamma(x)$ crosses over to a purely exponential decay, $\gamma(x)\sim e^{-x/\xi}$. In the intermediate regime of weak disorder, however, the decay is better captured as an exponential decay accompanied by a damped oscillation as
\bea
\label{eq:damp_oscillation}
\gamma(x) = A\cdot  e^{-x/\xi} + B\cdot \cos(2k_F x) \;e^{-x/\xi_O} \ ,
\eea
where $B$ is the oscillation amplitude and $\xi_O$ is the oscillation decay length. The fitting result shown in Fig.\ref{fig:attractive_1D}(b) for $W=t$ and $V=0$ yields $\xi=25.45$ and $\xi_O=4.00$, demonstrating that this fitting functional form accurately captures the decay behavior of $\gamma(x)$ in the weak-disorder regime. 

Building on the non-interacting benchmark results presented above, we now examine the effects of interactions. For repulsive interaction ($V>0$), Fig.\ref{fig:attractive_1D}(c) shows the result for $V=0.9t$ and $W=t$, where $\gamma(x)$ follows clear exponential decay with $\xi=8.99$ and $\xi_O \rightarrow0$, significantly shorter than the non-interacting value ($\xi=25.45$ and $\xi_O =4.00$) at the same disorder strength. This confirms that repulsive interactions enhance localization and also suppress oscillation. We note that the $\gamma(x)$ point at $x=1$ deviates from the exponential decay trend. We conjecture that this is due to the special form of the nearest-neighbor repulsion, which disfavors double occupation of adjacent sites; thus, we exclude it from the fitting and use the data from $x\ge2$.

Interestingly, for attractive interactions ($V<0$), we observe clear evidence of delocalization. As shown in Fig.~\ref{fig:attractive_1D}(d), for $V=-0.9t$ and $W=t$, the extracted localization length diverges to $\xi=456.86$ which is larger than the system size (with $\xi_O=1.32$), indicating an extended state. In this regime, $\gamma(x)$ remains well captured by the fitting form in Eq.~\ref{eq:damp_oscillation}, approaching a constant at large $x$. This result is in full agreement with previous numerical studies \cite{DMRG_1D_PhysRevB.72.024208}, which reported the same extended state under identical parameters. Altogether, these findings validate our theoretical analysis and demonstrate that the MDM framework, when combined with the DMRG algorithm, reliably extends to interacting systems and faithfully captures localization–delocalization behavior under both repulsive and attractive interactions. Nevertheless, an exact determination of the transition point in 1D remains challenging within our current approach and will be left for future investigation.

\subsection{2D Anderson-Hubbard model}
\label{sec:anderson_hubbard_interacting}

With the validity of our method established, we now address the central question of this section: whether metallic phases can exist in two dimensions under the combination of electron correlations and disorder. While previous quantum Monte Carlo studies have hinted at delocalization tendencies, the fermionic sign problem and the analytic continuation are still challenges in quantum Monte Carlo calculations \cite{DQMC_PhysRevLett.83.4610,DQMC_PhysRevLett.87.146401,DQMC_PhysRevLett.93.126401,DQMC_PhysRevB.84.035121}. 
On the other hand, our approach relies only on the ground-state wavefunction and quasi-1D scaling analysis. Ground states can be obtained with high accuracy, and the quasi-1D geometry is naturally compatible with DMRG simulations, making our method both more reliable and more efficient for tackling this problem.

To address this issue, we apply the many-body MDM approach to the Anderson-Hubbard model on a quasi-1D bar, which is described by the following Hamiltonian
\begin{equation}
\label{eq:anderson-hubbard}
H_{U} = -t \sum_{\langle (i,\alpha),(j,\beta)\rangle}\sum_{\sigma}
 c_{i,\alpha,\sigma}^\dagger c_{j,\beta,\sigma} +
U \sum_{(i,\alpha)} n_{i,\alpha,\uparrow} n_{i,\alpha,\downarrow}
+ \sum_{(i,\alpha),\sigma} 
\epsilon_{i,\alpha} n_{i,\alpha,\sigma}
\end{equation}
Here, $i,j$ and $\alpha,\beta$ follow the notation introduced earlier, $\sigma\in\{\uparrow,\downarrow\}$ denotes the spin index and $\epsilon_{i,\alpha}$ are uniformly distributed in $[-W,W]$.

\begin{figure}[htbp]
    \centering
    \includegraphics[width=1.02\linewidth]{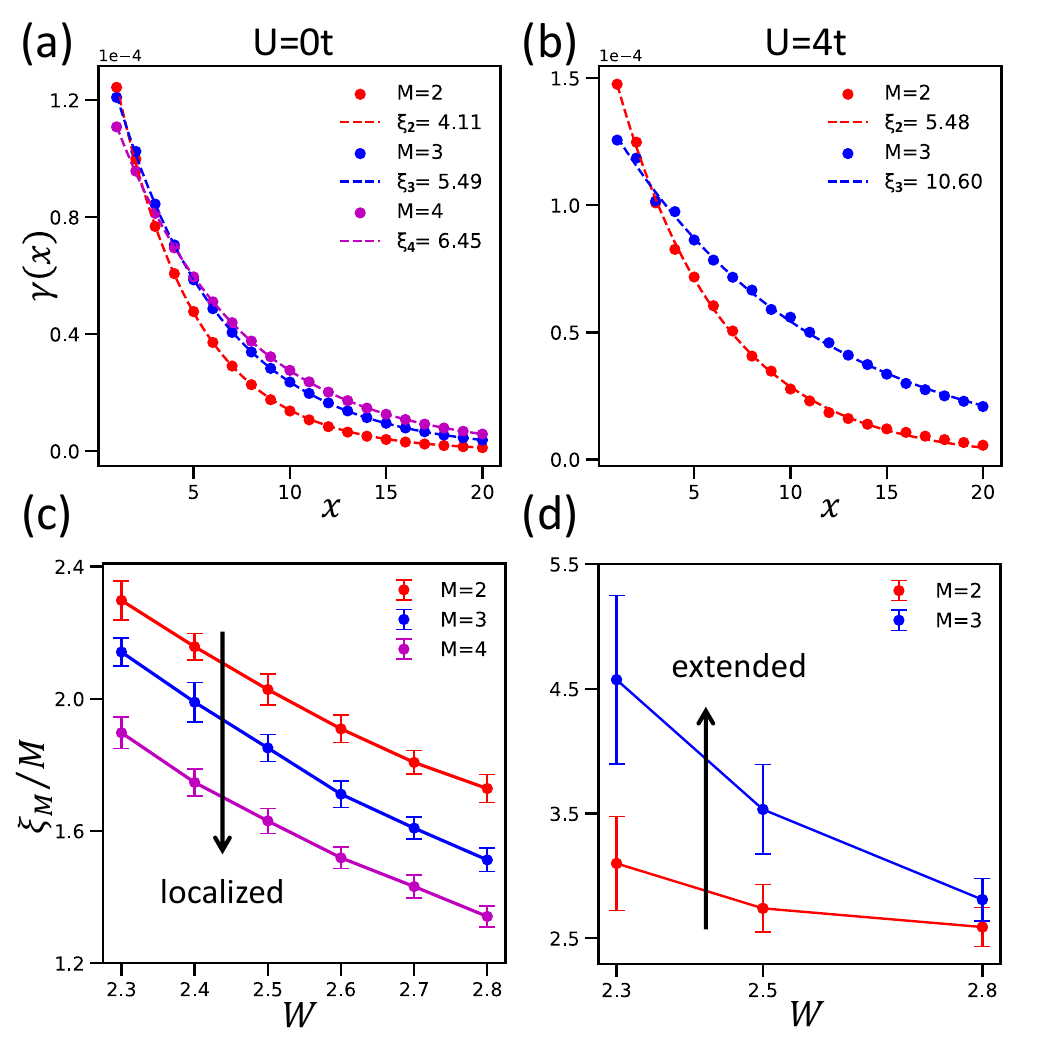}
    \caption{Results of $\gamma(x)$ and finite-size scaling of $\xi_M/M$ for the quasi-1D Anderson-Hubbard model. (a) shows the $\gamma(x)$ and exponential fitting results for $M=2,3,4$ in the non-interacting case with $U=0$ and $W=2.5t$, which is calculated by ED; (c) is the corresponding finite-size scaling of $\xi_M/M$ for $W=2.3t$--$2.8t$. (b) shows the $\gamma(x)$ and exponential fitting results for $M=2,3$ with $U=4t$ and $W=2.5t$, which is calculated by DMRG; (d) is the corresponding finite-size scaling of $\xi_M/M$ for $W=2.3t$--$2.8t$. All disorder realizations are on systems of size $M\times L$ with $L=60$ under OBC, and in the site averaging for each sample, the 10 sites nearest to each boundary are excluded to minimize open-boundary effects. In (c), data points are disorder averages over 2000 realizations, with error bars indicating standard errors estimated from 20 independent subsets. In (d), each point is averaged over 100 realizations, and error bars are standard errors obtained from five subsets of 20 samples.    }
    \label{fig:2d_correlation}
\end{figure}

It is well-known that the Hubbard model hosts a remarkably rich phase diagram with numerous competing orders, most prominently Mott physics at half-filling. In this work, our focus is on the correlated disordered metallic regime. To avoid interference from other ordered phases, we consider an electron filling of $n=4/15$, which is far away from half-filling. At this density, we expect the $U \rightarrow \infty$ limit corresponds to a correlated metal without other competing orders. The problem then reduces to the competition between a correlated metallic state and an Anderson insulator, providing a natural setting to investigate the delocalizing role of Hubbard interactions and the emergence of metallic behavior in disordered 2D systems.

We define the SDM for the Anderson-Hubbard model using the two ground states $\ket{GS_N}$ and $\ket{GS_{N-1}}$ as $\rho^{\mathrm{sub}}_{(i,\alpha);(j,\beta)}
= \sum_{\sigma}\Big(\langle c_{i,\alpha,\sigma}^\dagger c_{j,\beta,\sigma}\rangle_{N}-\langle c_{i,\alpha,\sigma}^\dagger c_{j,\beta,\sigma}\rangle_{N-1}\Big)$. Since the system preserves time-reversal symmetry, the SDM remains the same whether $\ket{GS_N}$ and $\ket{GS_{N-1}}$ differ by a spin-up or a spin-down fermion. Because DMRG calculations remain computationally demanding for $M\ge4$, our finite-U calculation to extract the localization lengths is restricted to bars of width $M<4$, which represents the practical computational limit of our study. 
For $M=2$, we perform 165 steps of sweep with the number of kept states increasing gradually to 2500, ensuring a truncation error $\epsilon \leq 10^{-8}$;  
for $M=3$, 195 steps of sweep are carried out with the number of kept states increasing gradually to 4000, achieving $\epsilon \leq 10^{-6}$.  
Further computational details and single-sample benchmarks with ED in the non-interacting case ($U=0$) are presented in Appendix \ref{supp:benchmark}.
For comparison, the localization lengths at $U=0$ are calculated by ED of the non-interacting Hamiltonians in the quasi-1D bars of widths up to $M=4$ under OBC. These system sizes are already sufficient to capture the finite-size scaling characteristics that distinguish metallic from insulating behavior, as discussed below.

We first examine the non-interacting case with $U=0$. Within the considered disorder range $W\in[2.3t,2.8t]$, $\gamma(x)$ exhibits robust exponential decay, $\gamma(x)\sim e^{-x/\xi_M}$, from which the localization lengths $\xi_M$ can be precisely extracted, as shown in Fig.\ref{fig:2d_correlation}(a) for the representative results at disorder strength $W=2.5t$. The corresponding finite-size scaling of $\xi_M/M$ is presented in Fig.~\ref{fig:2d_correlation}(c), which displays clear localized-state scaling behavior as expected. Specifically, for a fixed width $M$, $\xi_M/M$ decreases as $W$ increases; and for a fixed disorder strength $W$, $\xi_M/M$ decreases with increasing $M$, confirming the absence of metallic scaling in the non-interacting case \cite{scaling_PhysRevLett.42.673}.

We then turn to the interacting case with $U=4$. Remarkably, even in the presence of strong Hubbard interactions, $\gamma(x)$ retains a clean exponential decay form, as shown in Fig.\ref{fig:2d_correlation}(b) for the representative results at disorder strength $W=2.5t$. Additional numerical results of $\gamma(x)$ and the corresponding exponential fits at other disorder strengths can be found in Appendix \ref{supp:anderson-hubbard}.
Moreover, for the same disorder strength $W$ and width $M$, the localization length at $U=4t$ is significantly larger than that at $U=0t$, with the enhancement becoming more pronounced as $M$ increases. For example, we have $\xi_2(U=4t)/\xi_2(U=0t)\approx1.33$ while $\xi_3(U=4t)/\xi_3(U=0t)\approx1.93$ at disorder strength $W=2.5t$.
This directly impacts the finite-size scaling of $\xi_M/M$ in Fig.~\ref{fig:2d_correlation}(d), where $\xi_M$ still decreases with increasing $W$; but for fixed $W$, $\xi_M$ now increases with system width $M$.  Such scaling behavior is the hallmark of a 2D metallic phase.
%$\textcolor{red}{Uc??}$

These results demonstrate that finite disorder in the 2D Hubbard model can indeed host a correlated metallic state, in sharp contrast to the localized behavior of the non-interacting case. This finding provides direct numerical evidence that Hubbard interactions can play a delocalizing role in disordered 2D systems, establishing a new route to metallicity beyond the conventional Anderson paradigms.

\section{Summary and Outlook}
\label{sec:summary_and_outlook}
In this work, we develop a general framework for extracting localization lengths in disordered quantum systems employing the density matrix. By reformulating localization in terms of the modulus of the one-particle density matrix, our method (MDM) provides a direct and statistically robust way to determine localization lengths in quasi-1D geometries. Benchmarking against the standard transfer matrix method, we show that MDM faithfully reproduces the metal–insulator transition in the 3D Anderson model, captures the energy-dependent mobility edge, and remains consistent across different universality classes, including the 2D spin–orbit coupled case. We further demonstrated that MDM can be readily applied to multiorbital Anderson systems by benchmarking it against TMDCA for a two-orbital model. These results establish MDM as a reliable and computationally efficient alternative to TMM.

We then generalize the framework to interacting systems through the many-body subtraction density matrix. Using DMRG calculations, we demonstrated that this extension captures localization physics in disordered 1D spinless fermion chains with nearest-neighbor interactions, and reproduces the emergence of extended states at specific attractive interaction strength, consistent with previous numerical studies. Finally, we apply the method to the 2D Anderson–Hubbard model at $n=4/15$ filling, showing that Hubbard interactions can enhance delocalization and stabilize a correlated metallic phase at finite disorder in two dimensions. Together, these results establish the MDM framework as a versatile and powerful tool for exploring Anderson localization and its interplay with electronic correlations.

Looking ahead, several promising directions emerge from this work. First, integrating the MDM framework into broader computational approaches, such as density matrix embedding theory and density functional theory, could open powerful routes for investigating localization in realistic materials. Second, the method can be extended to other interacting systems and low-dimensional metal-insulator transitions, including low-density electron gases. Third, the MDM formalism can be naturally combined with other tensor-network algorithms \cite{RevModPhys.93.045003,xiang2023density} (e.g., projected entangled pair states), enabling studies of larger systems and providing new insights into entanglement and localization in many-body settings.
Taken together, our study demonstrates that the density matrix framework offers both conceptual simplicity and computational versatility, opening a new perspective on Anderson localization and its extensions to interacting and correlated systems.

\section{Acknowledgement}
We thank Jiale Huang for the useful discussion and help in DMRG. 
We acknowledge the support by the National Key R\&D Program of China (Grant No. 2022YFA1403800), the National Natural Science Foundation of China (No. NSFC-12174428, No. NSFC-12274279, and NSFC-12274290),  and the Chinese Academy of Sciences Project for Young Scientists in Basic Research (2022YSBR-048). H.W. also acknowledges support from the New Cornerstone Science Foundation through the XPLORER PRIZE.

\bibliography{reference}

\bibliographystyle{apsrev4-2}

\onecolumngrid

\newpage
\clearpage

\appendix

\section{ Comparative Frameworks for Localization Length: Transfer Matrix and Modular Density Matrix}
\label{supp:method_comparison}

In this section, we provide a systematic analysis and comparison of our MDM approach with TMM, in order to clarify the connections and differences in how the localization length is defined in these two methods. 
We first review the standard TMM, and then compare it with our newly introduced MDM approach.  The analysis highlights both the formal connections and the key differences. 
We further demonstrate mathematically the equivalence of the localization lengths obtained from the TMM and the MDM approach, which is consistent with our numerical results. 
For clarity, we denote 
$\xi_{TM}$ as the localization length obtained from the TMM, 
and $\xi_{MDM}$ as that from the MDM approach in this section.

\subsection{Transfer matrix method }
\label{supp:TMM_details}
We first provide a detailed description of TMM \cite{TMM_MacKinnon_1981,TMM_MacKinnon_1983,FSS_Pichard_1981} and give the definition of the localization length in  this method. Consider a quasi-1D system of width $M$ and length $L$, the Schrödinger equation for the quasi-1D system can be written as
\bea
\label{eq:slice_ham}
H_i \phi_i + V_{i,i+1} \phi_{i+1} + V_{i,i-1} \phi_{i-1} = E \phi_i
\eea
Here, $H_i$ denotes the block of Hamiltonian matrix elements for the $i$-th slice, 
$V_{i,i+1}$ represents the block of Hamiltonian matrix elements corresponding to the hopping terms between the $i$-th and $(i+1)$-th slices, 
and $\phi_i$ is the component of the eigen-wavefunction $\ket{\psi_E}$ with eigenenergy $E$ on the $i$-th slice, which can be expressed as $\phi_i = \ket{i}\braket{i|\psi_E}$. 
Based on Eq.~\ref{eq:slice_ham}, we then define the single transfer matrix $T_i(E)$ between adjacent slices as
\bea
\label{eq:single_tmm_define}
\left ( \begin{matrix}
 \phi_{i+1}\\\phi_{i}
\end{matrix} \right )=T_i(E)\left ( \begin{matrix}
 \phi_i\\\phi_{i-1}
\end{matrix} \right )\;, \; T_i(E)= \left ( \begin{matrix}
V^{-1}_{i,i+1}(E-H_i) & -V^{-1}_{i,i+1}V_{i,i-1} \\ I_{M^{d-1}\times M^{d-1}} & 0_{M^{d-1}\times M^{d-1}}
 \end{matrix} \right )
\eea
For a sysem with length $L$, the total transfer matrix $S_L(E)$ is given by the ordered product of the single transfer matrices $T_i(E)$, which reads
\bea
\label{eq:total_tmm_define}
S_L(E) = T_L(E)T_{L-1}(E)\cdots T_1(E)
\eea
According to the Furstenberg–Kesten law of large numbers \cite{RMT_furstenberg_1960} and Oseledec’s multiplicative ergodic theorem \cite{RMT_Crisanti_1993} for random matrix products, the localization length is defined in terms of the smallest positive Lyapunov exponent as
\bea
\xi_{TM}=\frac{1}{\gamma_1} \;,\; \gamma_1=\lim_{L\rightarrow\infty} \frac{1}{2L}ln(|| S^\dagger_L(E) S_L(E) ||)
\eea
The Lyapunov exponent can be obtained stably through iterative QR or SVD decompositions \cite{FSS_Pichard_1981}. A detailed analysis of the statistical stability of the localization length $\xi_{TM}$ with increasing length $L$ will be presented later, here we first provide its definition within the TMM framework.

\subsection{Relation between TMM and MDM}

The definitions of the localization length from MDM and TMM introduced above 
can be placed on a common footing by analyzing them from the perspective of single-particle wavefunctions and their associated density matrices. 
To clarify the relationships between the localization lengths defined from these two methods, we focus on the 1D case. 
The conclusions drawn from this analysis also hold for quasi-1D systems by simply adding the internal degrees of freedom within a slice.

In the TMM framework, according to Eq.\ref{eq:single_tmm_define} and Eq.\ref{eq:total_tmm_define}, 
the eigen-wavefunction component at the $x$-th site propagated from the reference $0$-th site can be written as
\bea
\Phi_x=S_x(E)\;\Phi_0\;,\;\Phi_x =\left ( \begin{matrix}
 \phi_{x+1}\\\phi_{x}
\end{matrix} \right )
\eea
Where $\phi_x$ is the component of the eigen-wavefunction $\ket{\psi_E}$ with eigenenergy $E$ at the $x$-th site, which can be expressed as $\phi_x=\braket{x|\psi_E}=\psi_E(x)$ (follow the same convention as in Eq.\ref{eq:slice_ham}). Then the Euclidean norm of $S_x(E)\Phi_0$ can be expressed in the form of $\psi_E$ as
\bea
||S_x \Phi_0||=||\Phi_x||=\sqrt{|\psi_E(x)|^2+|\psi_E(x+1)|^2}\approx \sqrt{2}\cdot |\psi_E(x)|
\eea
Combining this with the prerequisite that the wavefunction component at the reference $0$-th site is set to unity in TMM, $|\psi_E(0)|=1$, the Euclidean norm can be rewritten as follow
\bea
\label{eq:tm_mdm_relation}
||S_x \Phi_0||\approx \sqrt{2}\cdot|\psi_E(x)|\cdot|\psi_E(0)|  \propto|\bra{\psi_E}c_0^\dagger c_x \ket{\psi_E}| 
\eea
This is precisely the 1D form of the basic building block of the MDM in Eq.\ref{eq:MDM_definition} in the main text, demonstrating the direct correspondence between the TMM and the MDM.

Up to this point, from the perspective of the eigen-wavefunction and the MDM, we have established that the total transfer-matrix quantity $||S_x(E)||$ is directly correlated with the basic building block of the MDM (Eq.\ref{eq:MDM_definition}). Nevertheless, the MDM definition of localization length focuses on the same building block $|\bra{\psi_E}c_0^\dagger c_x \ket{\psi_E}| $ but employs a different averaging scheme. Instead of taking the limit of $x\rightarrow\infty$, one considers the averaged quantity $\sum_{i}|\bra{\psi_E}c_i^\dagger c_{i+x} \ket{\psi_E}|$, which collects contributions from all sites separated by distance $x$. The slowest decaying mode $\gamma(x)$ in the MDM is then fitted to $e^{-x/\xi_{MDM}}$ to obtain the localization length $\xi_{MDM}$. This is the core difference between the MDM and TMM framework.

\subsection{The equivalence and statistical stability of $\xi_{TM}$ and $\xi_{MDM}$}

Based on the two definitions of the localization length introduced above, we now provide a mathematical demonstration of their numerical stability after statistical averaging, as well as their equivalence in the localized regime.

We first demonstrate the numerical stability of the localization length $\xi_{TM}$ obtained from the TMM. 
For convenience, we denote the logarithm of the product transfer matrix norm as 
$U_x(E) \coloneqq \ln(\|S_x(E)\Phi_0\|)$. 
The Furstenberg–Kesten law of large numbers \cite{RMT_furstenberg_1960} and Oseledec’s theorem \cite{RMT_Crisanti_1993} then imply
\bea
\label{eq:ux_x_infinite}
\lim_{x\rightarrow\infty}\frac{U_x}{x}\rightarrow\gamma_1, \;\xi_{TM}=\frac{1}{\gamma_1}
\eea
Here, $\gamma_1$ is the largest positive Lyapunov exponent of $S_x(E)$, which is also named as the first Lyapunov cumulant, representing the mean value of $\frac{U_x}{x}$.

Beyond the mean value, the central limit theorem for random matrix products \cite{RMT_Page_1982,RMT_Crisanti_1993,RMT_Beenakker_1997} yields a Gaussian distribution of $U_x$ with finite $x$,  which is
\bea
\label{eq:lya_12}
\frac{U_x-\gamma_1 x}{\sqrt{x}} \sim \mathcal{G}(0,\gamma_2) 
\eea
Here, $\mathcal{G}(\mu,\sigma^2)$ denotes the Gaussian distribution with the mean value $\mu$ and variance $\sigma^2$. $\gamma_2$ is named as the second Lyapunov cumulant, representing the variance of $\frac{U_x-\gamma_1 x}{\sqrt{x}}$. Equivalently, Eq.\ref{eq:lya_12} can be rewritten as the distribution of $\frac{U_x}{x}$ with finite $x$, which reads
\bea
\label{eq:ux_x_finite}
\frac{U_x}{x}\sim \mathcal{G}(\gamma_1,\frac{\gamma_2}{x})
\eea
Therefore, as $x\rightarrow\infty$, Eq.\ref{eq:ux_x_finite} reduces to Eq.\ref{eq:ux_x_infinite} with $\frac{\gamma_2}{x}\rightarrow0$, yielding a stable localization length $\frac{U_x}{x}=\frac{1}{\gamma_1}=\xi_{TM}$.

Nevertheless, in the MDM framework we consider $\rho^m(x)$ at finite $x$, and thus the variance $\gamma_2$ cannot be neglected. 
After statistical averaging, however, a stable exponential decay can still be obtained in the localized regime. 
The exponentially decaying mode $\gamma(x)$ (defined in Eq.~\ref{eq:symmetrization}) extracted from the MDM satisfies
$
\gamma(x) = |\rho^m(x)|^2 \propto \langle \big|\bra{\psi_E}c_0^\dagger c_{x} \ket{\psi_E}\big|\rangle^2  \propto \langle e^{-U_x}\rangle^2
$.
According to Eq.\ref{eq:ux_x_finite}, one has $U_x \sim \mathcal{G}(\gamma_1 x,\gamma_2 x)$. 
Therefore, averaging over samples and sites in Eq.~\ref{eq:MDM_definition} is equivalent to evaluating $\left\langle e^{-U_x} \right\rangle^2$ analytically, which can be derived from Gaussian integration as
\bea
\gamma(x)=|\rho^m(x)|^2 \propto \left \langle  e^{-U_x}\right \rangle^2= \Big(\int_{-\infty}^{+\infty} dU_x \;e^{-U_x} \frac{1}{\sqrt{2\pi\gamma_2 x}} e^{-\frac{(U_x-\gamma_1 x)^2}{2\gamma_2 x}}    \Big)^2=A \cdot e^{-(2\gamma_1 -\gamma_2) x}
\eea
Here $A$ is a constant coefficients obtained from integration, and the asymptotic decay of MDM is determined by both $\gamma_1$ and $\gamma_2$ in general.

Notably, previous studies \cite{RMT_Beenakker_1997} have established that,
for 1D and quasi-1D systems in the localized regime with finite-variance, short-range disorder, the mean value $\gamma_1$ and the variance $\gamma_2$ of the logarithmic wavefunction amplitude are equal, so that
\bea
\label{eq:gamma1_gamma2}
\gamma_1=\gamma_2
\eea
This identity is a hallmark of single-parameter scaling in the localized regime and underlies the log-normal distribution of transmission. 
The non-interacting systems considered in our work all lie in the localized regime with finite-variance short-range disorder, where single-parameter scaling applies and Eq.\ref{eq:gamma1_gamma2} is satisfied. 
Therefore, the localization length defined via the MDM is theoretically identical to that obtained from the TMM, satisfying 
\bea
\xi_{MDM} = \xi_{TM} = \tfrac{1}{\gamma_1}.
\eea
This analytical result is fully consistent with our numerical calculations in Fig.\ref{fig:TM_DM_compare} and Appendix \ref{Supplement:non-interaction}.

\section{ Supplementary results of non-interacting systems \label{Supplement:non-interaction}}

In this section, we provide supplementary details and results for the models discussed in the main text, including the single-orbital 3D Anderson model and the spinful 2D Anderson model with random SOC.
%, and the two-orbital 3D Anderson model with intra-orbital binary disorder. 
For the 2D case, we further compare the scaling behavior of the Anderson model with and without SOC. 
This comparison clearly demonstrates that, in the absence of SOC, all states remain localized for any finite disorder, whereas the presence of SOC places the system in the symplectic universality class, allowing the emergence of a metallic phase at finite disorder and thereby leading to a finite critical disorder strength $W_c$ for metal-insulator transition. All results presented in this section confirm that the localization length extracted from the MDM quantitatively reproduces that from the TMM.
These additional results further establish the accuracy and general applicability of the MDM framework to Anderson localization across different dimensions and symmetry classes.

\subsection{3D Anderson model } 
\label{supp:3D-anderson}

\begin{figure}[H]
    \centering
    \includegraphics[width=1.0\linewidth]{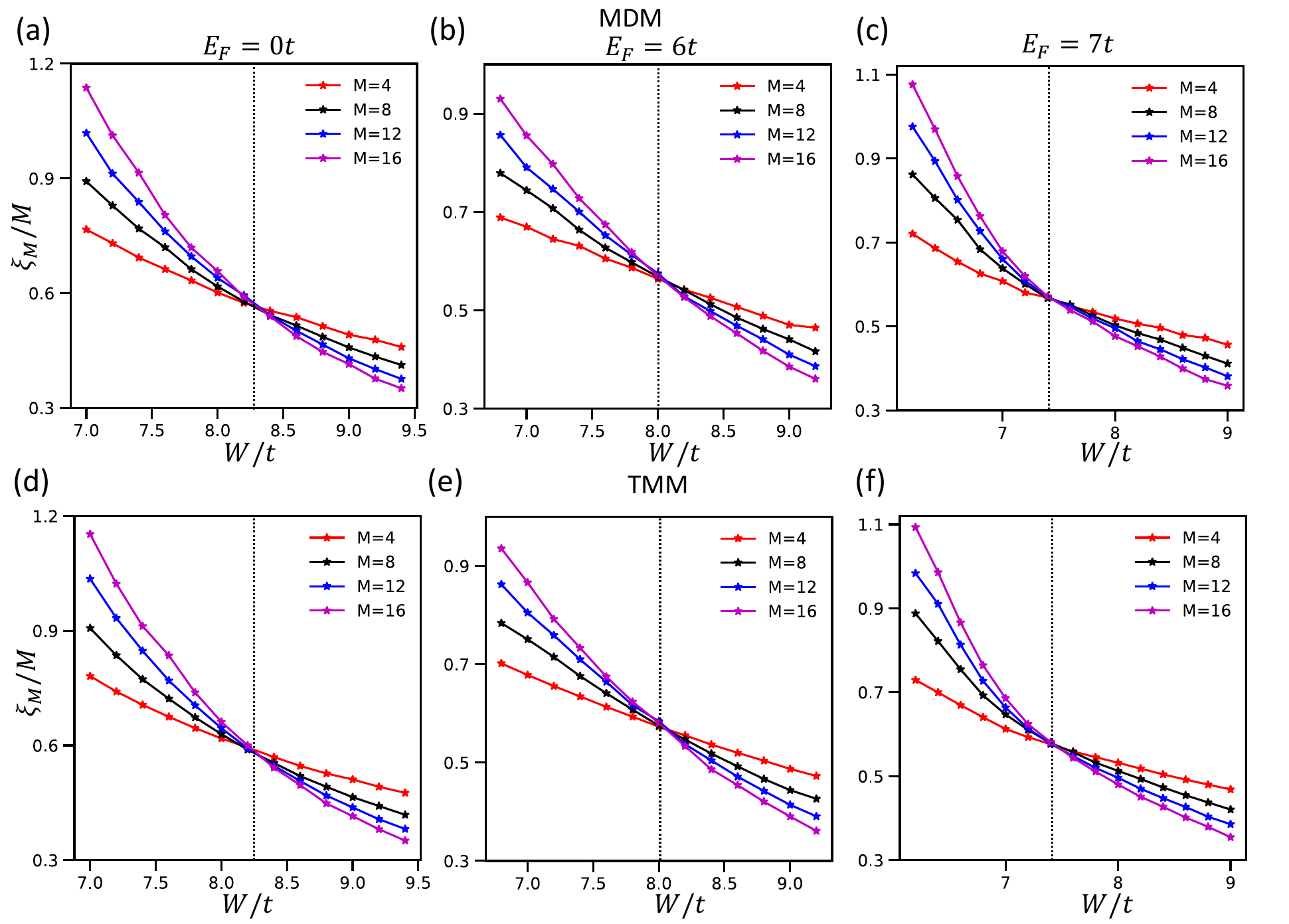}
    \caption{ Comparison of localization length scaling between MDM and TMM for the 3D Anderson model. The plots show $\xi_M/M$ as a function of disorder strength $W$ at fixed Fermi energies $E_F=0,6,7$ for different quasi-1D widths $M$. (a)-(c) present MDM results, and (d)-(f) show the corresponding TMM results. The critical disorder strengths are $W_c\approx 8.3, \;8.0,\; 7.4$ for $E_F=0,\;6,\;7$, respectively. Each $\xi_M/M$ data point from the MDM is obtained by averaging 3000 disorder realizations of size $M^2\times L$ with $L=300$, while in the TMM each data is obtained from quasi-1D systems of length $L=3\times10^6$ by averaging three disorder realizations. }
    \label{fig:3D_DM_EF_fix}
\end{figure}

\begin{figure}[H]
    \centering
    \includegraphics[width=1.0\linewidth]{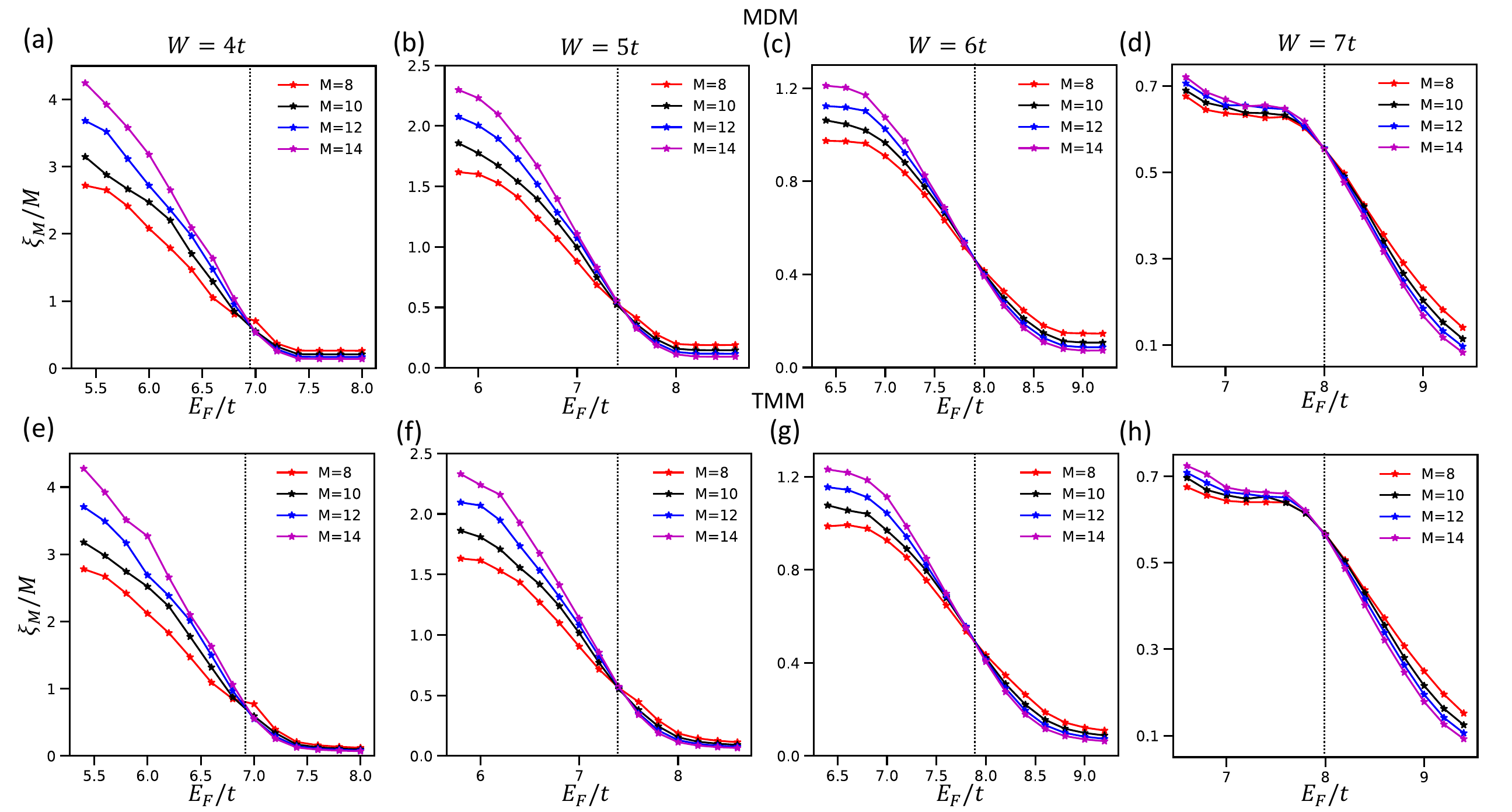}
    \caption{ Comparison of localization length scaling between MDM and TMM for the 3D Anderson model. The plots show $\xi_M/M$ as a function of Fermi energy $E_F$ at fixed disorder strengths $W=4,5,6,7$ for different quasi-1D widths $M$. (a)-(d) present MDM results, and (e)-(h) show the corresponding TMM results. The critical Fermi energies are $E_{F_c}\approx 6.9, \;7.4,\; 7.9,\;8.0$ for $W=4,\;5,\;6,\;7$, respectively. Each $\xi_M/M$ data point from the MDM is obtained by averaging 3000 disorder realizations of size $M^2\times L$ with $L=300$, while in the TMM each data is obtained from quasi-1D systems of length $L=3\times10^6$ by averaging three disorder realizations.}
    \label{fig:3D_DM_W_fix}
\end{figure}

The single-orbital 3D Anderson model on a quasi-1D cubic lattice has been defined in the main text (Eq.~\ref{eq:anderson_model}). 
The nearest-neighbor hopping is set to $t=1$ as the energy unit, and onsite random potentials $\epsilon_i$ are uniformly distributed within $[-W,W]$. 
Fig.\ref{fig:3D_DM_EF_fix} compares the localization lengths obtained from the MDM and TMM at different fixed Fermi energies $E_F$ as a function of disorder strength $W$. 
As $E_F$ deviates from the band center, the critical disorder strength $W_c$ decreases slowly. 
The transition point $W_c$ separates two regimes: in the metallic phase, $\xi_M/M$ increases with $M$ at fixed $W$, whereas in the Anderson insulating phase, $\xi_M/M$ decreases with $M$. 
Fig.\ref{fig:3D_DM_W_fix} presents the complementary analysis at fixed $W$ as a function of $E_F$. 
Here, the mobility edge separating extended and localized states is clearly visible. 
With increasing $W$, the critical Fermi energy $E_{F_c}$ shifts upward, reflecting the characteristic mobility-edge structure of the 3D Anderson model. 
In both Fig.\ref{fig:3D_DM_EF_fix} and Fig.\ref{fig:3D_DM_W_fix}, the normalized localization length $\xi_M/M$ and the MIT critical point $W_c$ ($E_{F_c}$) extracted from the MDM agree precisely with those obtained from the TMM, thereby demonstrating the validity and accuracy of the MDM approach across the entire phase space of the 3D Anderson model.
                                                                                                   
\subsection{Spinful 2D Anderson model with SOC} 
\label{supp:2D-anderson-SOC}

The spinful 2D Anderson model with SOC \cite{SOC_Asada_2002} on a quasi-1D square lattice has been defined in the main text (Eq.\ref{eq:2D_SOC}). In this model, both the onsite potentials and the nearest-neighbor SOC terms are random. The onsite disorder $\epsilon_i$ is uniformly distributed in the interval $[-W,W]$, the nearest-neighbor hopping incorporates random spin rotations described  by an SU(2) matrix $R_{(i,\alpha);(j,\beta)}$ acting on the spin space, which is written as 
\bea
R_{(i,\alpha);(j,\beta)}=\begin{pmatrix}
e^{ia_{(i,\alpha);(j,\beta)}} \cos b_{(i,\alpha);(j,\beta)} & e^{ic_{(i,\alpha);(j,\beta)}} \sin b_{(i,\alpha);(j,\beta)} \\
- e^{-ic_{(i,\alpha);(j,\beta)}} \sin  b_{(i,\alpha);(j,\beta)} & e^{-ia_{(i,\alpha);(j,\beta)}} \cos b_{(i,\alpha);(j,\beta)}
\end{pmatrix}
\eea
Here, $i$ and $j$ index slices along the longitudinal direction of the quasi-1D bar, and $\alpha,\beta$ denote internal site degree of freedom within a slice. $a_{(i,\alpha);(j,\beta)}$ and $c_{(i,\alpha);(j,\beta)}$ are uniformly distributed in $[0,2\pi)$, while $b_{(i,\alpha);(j,\beta)}$ are distributed with probability density $P(b)db=sin(2b)db$ for $0\le b\le 2\pi$. These distributions ensure that $R_{(i,\alpha);(j,\beta)}$ is uniformly distributed with respect to the Haar measure on SU(2) \cite{SOC_Asada_2002}. Hopping matrices on different bonds are statistically independent.

\begin{figure}[H]
    \centering
    \includegraphics[width=1.0\linewidth]{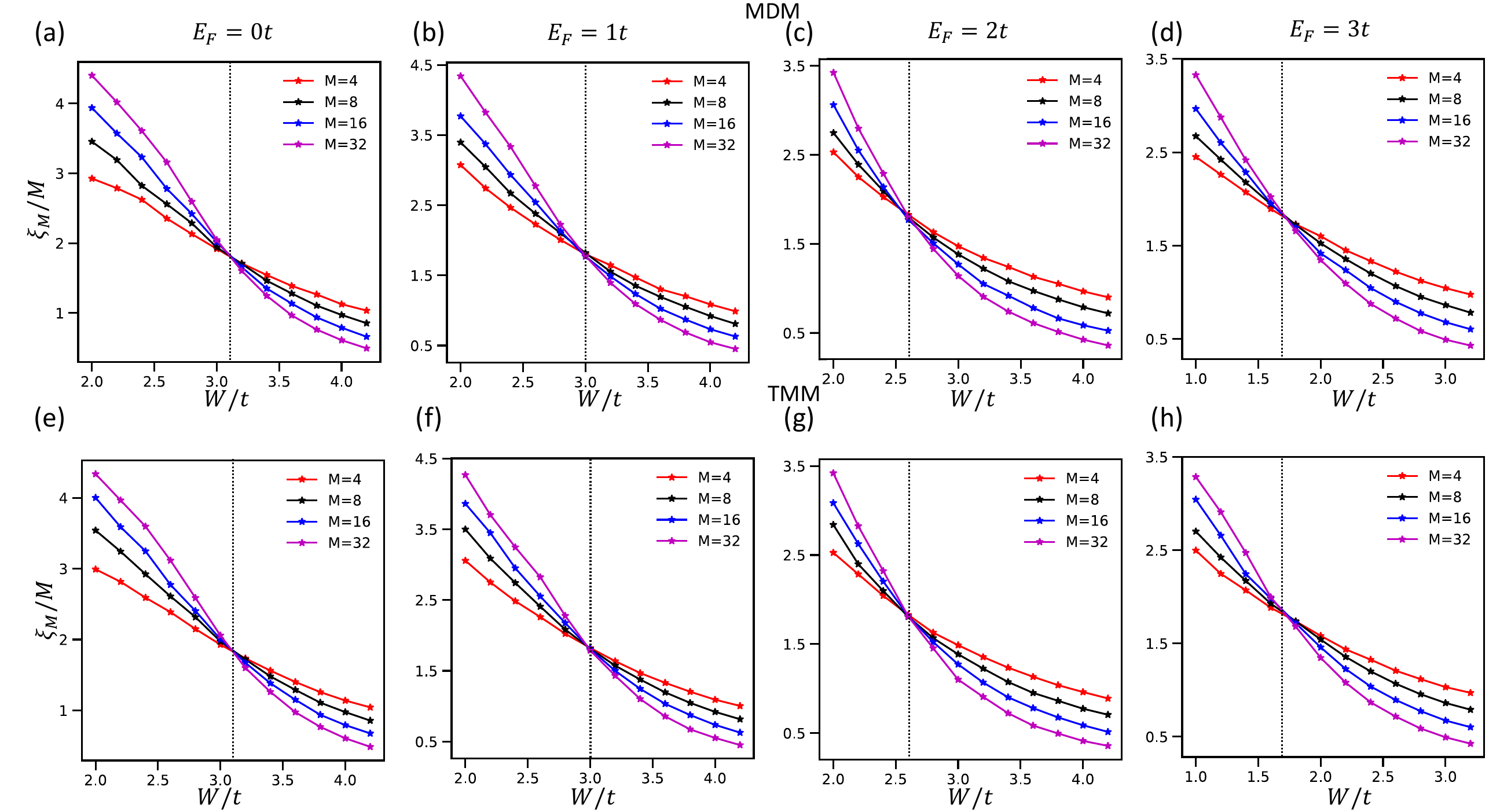}
    \caption{ Comparison of localization length scaling between MDM and TMM for the spinful 2D Anderson model with SOC. The plots show $\xi_M/M$ as a function of disorder strength $W$ at fixed Fermi energies $E_F=0,1,2,3$ for different quasi-1D width $M$. (a)-(d) present MDM results, and (e)-(h) show the corresponding TMM results. The critical disorder strengths are $W_c\approx 3.1, \;3.0,\; 2.6,\;1.7$ for $E_F=0,\;1,\;2,\;3$, respectively. Each $\xi_M/M$ data point from the MDM is obtained by averaging 3000 disorder realizations of size $M\times L$ with $L=500$, while in the TMM each data point is obtained from quasi-1D systems of length $L=3\times10^6$ by averaging three disorder realizations. }
    \label{fig:2D_SOC_DM_EF_fix}
\end{figure}

Supplementary localization scaling results from MDM and TMM are shown in Fig.\ref{fig:2D_SOC_DM_EF_fix}. These plots display the scaling of $\xi_M/M$ at fixed Fermi energies as a function of disorder strength $W$. 
As the Fermi energy deviates from $E_F=0$,  the MIT critical point $W_c$ gradually decreases. For each system width $M$, the normalized localization length $\xi_M/M$ decreases monotonically with increasing disorder. The critical point $W_c$ separates the metallic phase, where $\xi_M/M$ increases with $M$, from the Anderson insulating phase, where $\xi_M/M$ decreases with $M$.

\begin{figure}[H]
    \centering
    \includegraphics[width=0.9\linewidth]{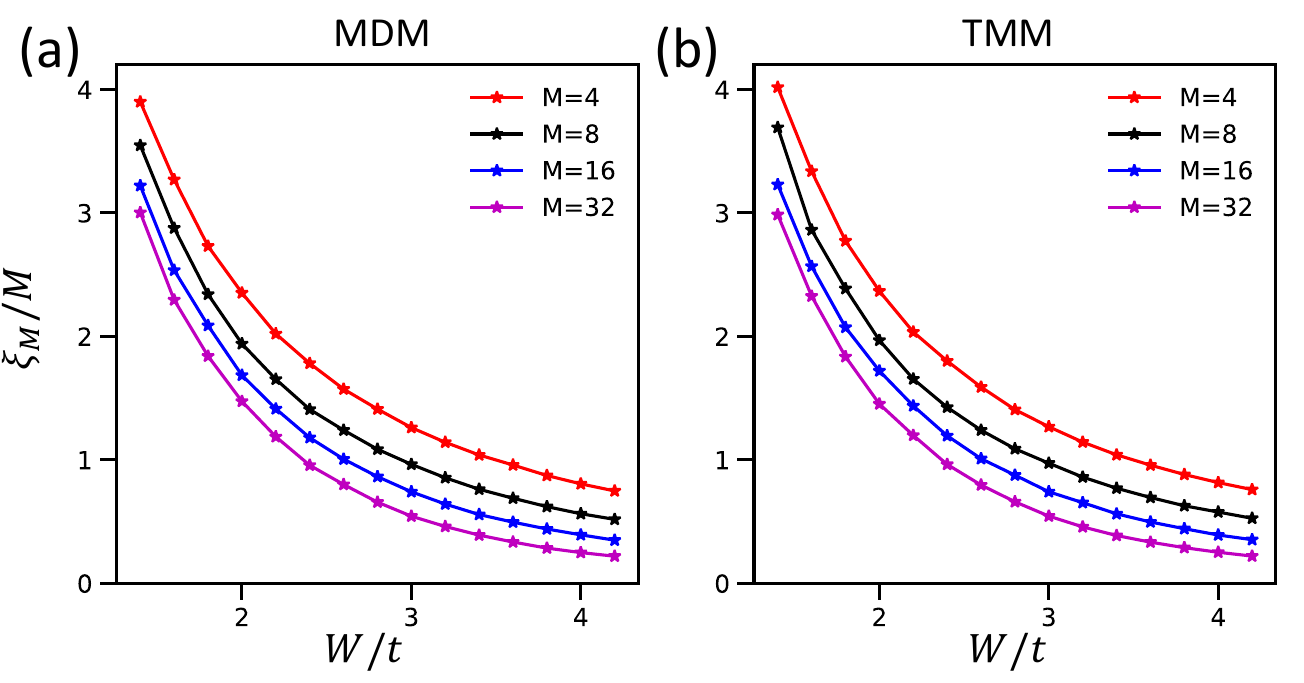}
    \caption{ Comparison of localization length scaling between MDM and TMM for the  2D Anderson model without SOC. The plots show $\xi_M/M$ as a function of disorder strength $W$ at fixed Fermi energy $E_F=0$ for different quasi-1D width $M$. (a) presents MDM result, and (b) shows the corresponding TMM result. Each $\xi_M/M$ data point from the MDM is obtained by averaging 3000 disorder realizations of size $M\times L$ with $L=500$, while in the TMM each data point is obtained from quasi-1D systems of length $L=3\times10^6$ by averaging three disorder realizations. }
    \label{fig:2D_EF0_nonint}
\end{figure}

To highlight the role of symmetry, we also compute the scaling behavior of the 2D Anderson model without SOC (taking the same Hamiltonian form as Eq.\ref{eq:anderson_model}), which belongs to the orthogonal class. As shown in Fig.\ref{fig:2D_EF0_nonint}, both MDM and TMM confirm that in this case all states remain localized for finite disorder, with no metallic phase \cite{SOC_10.1143/PTP.63.707}. In contrast, with SOC the system belongs to the symplectic class and supports a finite-disorder metallic phase \cite{SOC_10.1143/PTP.63.707,BERGMANN19841_soc,symmetry_PhysRevB.55.1142,ALT_Evers_2008,efetov1999supersymmetry}. In both cases (with and without SOC in 2D), the localization lengths $\xi_M/M$ obtained from MDM are in quantitative agreement with those from TMM. For the SU(2) model with SOC the critical disorder strengths $W_c$ extracted from two methods also precisely coincide. These results demonstrate the accuracy and reliability of the MDM approach across different universality classes.

\section{Physical Meaning of Subtraction Density Matrix}
\label{supp:spinful_SDM}
In the main text, we have discussed the relation of SDM to the operator $\hat{\psi}$ connecting $|GS_{N}\rangle$ and $|GS_{N-1}\rangle$. In this section, we want to have a general discussion on SDM. 
Here, we focus on the spinful version of SDM without losing generality, defined as  
\begin{equation}
\rho_{(i,\alpha);(j,\beta)}^{\mathrm{sub}}
=\sum_{\sigma=\uparrow,\downarrow}\Big(
\langle GS_{N}|c_{i,\alpha,\sigma}^\dagger c_{j,\beta,\sigma}|GS_{N}\rangle
-\langle GS_{N-1}|c_{i,\alpha,\sigma}^\dagger c_{j,\beta,\sigma}|GS_{N-1}\rangle
\Big)
\end{equation}
where $i,j$ index slices along the longitudinal direction of the quasi-1D bar, and $\alpha,\beta$ label the site degree of freedom within each slice (see Fig.\ref{fig:schematic_diagram}(c)). The states $|GS_{N}\rangle$ and $|GS_{N-1}\rangle$ are the many-body ground states with fixed particle numbers $N$ and $N-1$, respectively.

In the special case where $|GS_{N}\rangle = \ket{\psi_{E_F}} \otimes |GS_{N-1}\rangle$, corresponding to the non-interacting limit, 
the orthogonality of wavefunctions leads to the exact relation  $\rho_{(i,\alpha);(j,\beta)}^{\mathrm{sub}}
=\sum_{\sigma}\langle \psi_{E_F}|c_{i,\alpha,\sigma}^\dagger c_{j,\beta,\sigma}|\psi_{E_F}\rangle$, 
which reduces to the same form as the basic building block of MDM in the non-interacting case.

Beyond this simple product-state assumption, we consider the more general case where $|GS_{N}\rangle$ and $|GS_{N-1}\rangle$ are connected by an operator $\hat{\psi}$ such that $|GS_{N}\rangle=\hat{\psi}^\dagger|GS_{N-1}\rangle$. 
In the main text, we have already discussed the case $\hat{\psi}$ that comes from the linear combination of $c_{i\sigma}$.
For the Hubbard model or other interacting systems, $\hat{\psi}$ can also contain high-order terms,  such as $c_{i,\alpha,\sigma}n_{i,\alpha,\bar{\sigma}}$.
Therefore, we consider a more general form of the excitation operator,
\begin{equation}
\hat{\psi}=
\sum_{i,\alpha,\sigma} a^{(0)}_{i,\alpha,\sigma}\,c_{i,\alpha,\sigma}
+ a^{(1)}_{i,\alpha,\sigma}\,c_{i,\alpha,\sigma}n_{i,\alpha,\bar{\sigma}}
+\cdots=\sum_{n}\hat\psi^{(n)}
\end{equation}
Here, $a^{(0)}_{i,\alpha,\sigma}$ represents the amplitude of a bare fermionic excitation, while $a^{(1)}_{i,\alpha,\sigma}$ captures the contribution of a density-projected excitation involving the opposite spin $\bar{\sigma}$. $\hat{\psi}^{(n)}$ denotes the $n$-th order component of $\hat\psi$, e.g., $\hat{\psi}^{(0)}=\sum a^{(0)}_{i,\alpha,\sigma} c_{i,\alpha,\sigma}$ and $\hat{\psi}^{(1)}=\sum a^{(1)}_{i,\alpha,\sigma} c_{i,\alpha,\sigma}n_{i,\alpha,\bar{\sigma}}$. Higher-order terms $\hat\psi^{(n)}$ ($n\ge2$) may further encode multi-particle correlations in strongly interacting systems.

Based on the expansion of the excitation operator $\hat{\psi}$ introduced above, the SDM can be correspondingly expanded as  
\begin{equation}
\label{eq:total_rho_spinful}
\rho_{(i,\alpha);(j,\beta)}^{\mathrm{sub}}
=\sum_{n,m}\rho_{(i,\alpha);(j,\beta)}^{\mathrm{sub},(n,m)}-\sum_\sigma\langle GS_{N-1}|c_{i,\alpha,\sigma}^\dagger c_{j,\beta,\sigma}|GS_{N-1}\rangle
\end{equation}
where the superscript $(n,m)$ denotes the $(n,m)$-th order contribution associated with the corresponding term in $\hat{\psi}$. Explicitly, the $(n,m)$-th order component of SDM is given by  
\begin{equation}
\label{eq:rho_n}
\rho_{(i,\alpha);(j,\beta)}^{\mathrm{sub},(n,m)}
=\sum_{\sigma}
\langle GS_{N-1}|\hat{\psi}^{(n)} c_{i,\alpha,\sigma}^\dagger c_{j,\beta,\sigma} \hat{\psi}^{(m)\dagger}|GS_{N-1}\rangle
\end{equation}
Despite that the operator $\hat\psi$ may contain higher-order terms that lead to complicated algebraic relations, 
we postulate an effective fermion operator ansatz: without specifying the explicit form of higher-order corrections, 
we assume that the overall operator satisfies the canonical anticommutation relation $\{\hat\psi,\hat\psi^\dagger\}=1$. 
As a consistency check, we evaluate its expectation value in the many-body ground state $\ket{GS_{N-1}}$ as $\langle GS_{N-1}|\{\hat\psi,\hat\psi^\dagger\}|GS_{N-1}\rangle=\langle GS_{N}|GS_{N}\rangle +\langle GS_{N}|\hat\psi^\dagger\hat\psi^\dagger\hat\psi\hat\psi|GS_{N}\rangle  = 1+0=0$, 
which suggests that the effective normalization is approximately preserved within the low-energy subspace relevant for our construction.  
This provides a tentative theoretical justification for employing $\hat\psi$ as an effective fermionic excitation operator in the interacting case.

Using the anticommutation relation of $\hat\psi$, Eq.\ref{eq:total_rho_spinful} can be rewritten as 
\bea
\rho_{(i,\alpha);(j,\beta)}^{\mathrm{sub}}
&=\sum_{n,m}\langle\Big(\{\hat{\psi}^{(n)}, c_{i,\alpha,\sigma}^\dagger\}-c_{i,\alpha,\sigma}^\dagger\hat{\psi}^{(n)}\Big) \Big(\{c_{j,\beta,\sigma}, \hat{\psi}^{(m)\dagger}\}-\hat{\psi}^{(m)\dagger}c_{j,\beta,\sigma}\Big)\rangle_{N-1}-\langle c_{i,\alpha,\sigma}^\dagger c_{j,\beta,\sigma}\rangle_{N-1}\\
&=\sum_{n,m}\sum_{\sigma}\Big(\langle \{\hat{\psi}^{(n)}, c_{i,\alpha,\sigma}^\dagger\}\{c_{j,\beta,\sigma}, \hat{\psi}^{(m)\dagger}\}\rangle_{N-1}-
\langle \{\hat{\psi}^{(n)}, c_{i,\alpha,\sigma}^\dagger\} \hat{\psi}^{(m)\dagger}c_{j,\beta,\sigma}  \rangle_{N-1}
-\langle c_{i,\alpha,\sigma}^\dagger\hat{\psi}^{(n)}\{c_{j,\beta,\sigma}, \hat{\psi}^{(m)\dagger}\}\rangle_{N-1}\Big)
+\sum_{\sigma}\langle  c_{i,\alpha,\sigma}^\dagger (\hat\psi\hat\psi^\dagger-1) c_{j,\beta,\sigma}\rangle_{N-1}\\
&=\sum_{n,m}\sum_{\sigma}\Big(\langle \{\hat{\psi}^{(n)}, c_{i,\alpha,\sigma}^\dagger\}\{c_{j,\beta,\sigma}, \hat{\psi}^{(m)\dagger}\}\rangle_{N-1}-
\langle \{\hat{\psi}^{(n)}, c_{i,\alpha,\sigma}^\dagger\} \hat{\psi}^{(m)\dagger}c_{j,\beta,\sigma}  \rangle_{N-1}
-\langle c_{i,\alpha,\sigma}^\dagger\hat{\psi}^{(n)}\{c_{j,\beta,\sigma}, \hat{\psi}^{(m)\dagger}\}\rangle_{N-1}\Big)
+\sum_{\sigma}\langle  c_{i,\alpha,\sigma}^\dagger \hat\psi^\dagger\hat\psi c_{j,\beta,\sigma}\rangle_{N-1}\\
&\equiv\sum_{n,m}\chi_{(i,\alpha);(j,\beta)}^{\mathrm{sub},(n,m)}+\sum_{\sigma}\langle  c_{i,\alpha,\sigma}^\dagger \hat\psi^\dagger\hat\psi c_{j,\beta,\sigma}\rangle_{N-1}
\eea
Here, we use $\chi_{(i,\alpha);(j,\beta)}^{\mathrm{sub},(n,m)}$ to denote above long expressions, and $\langle \cdots\rangle_{N-1}$ is the simplified notation of $\bra{GS_{N-1}}\cdots\ket{GS_{N-1}}$. Despite the complexity of these expression, we can further simplify it as follows:
\begin{itemize}
\item First, for the case $n=m=0$, corresponding to the contribution of the bare fermionic excitation operator to the SDM, according to the anticommutation relations of $\hat\psi^{(0)}$ as follow 
\begin{equation}
\label{eq:anti_psi0}
\{\hat{\psi}^{(0)},c_{i,\alpha,\sigma}^\dagger\} = a_{i,\alpha,\sigma}^{(0)} \quad \;\;\quad \{\hat{\psi}^{(0)\dagger},c_{j,\beta,\sigma}\} = a_{j,\beta,\sigma}^{(0)*}
\end{equation}
one can obtain
\begin{equation}
\label{eq:chi_00}
\chi_{(i,\alpha);(j,\beta)}^{\mathrm{sub},(0,0)}=\sum_{\sigma} \Big(a^{(0)}_{i,\alpha,\sigma}\,a^{(0)\,*}_{j,\beta,\sigma}-a^{(0)}_{i,\alpha,\sigma}\langle\hat{\psi}^{(0)\dagger}c_{j,\beta,\sigma}\rangle_{N-1}-a^{(0)*}_{j,\beta,\sigma}\langle c_{i,\alpha,\sigma}^\dagger\hat{\psi}^{(0)}\rangle_{N-1}\Big)
\end{equation}
Here, $\sum_{\sigma} a^{(0)}_{i,\alpha,\sigma}\,a^{(0)\,*}_{j,\beta,\sigma}
=\sum_{\sigma}\,\langle \psi^{(0)}\,|\,c_{i,\alpha,\sigma}^\dagger\,c_{j,\beta,\sigma}\,|\,\psi^{(0)}\rangle$ encode the localization properties of $\hat\psi^{(0)}$, which follows the same expression as in Eq.\ref{eq:SDM_derivation} in the main text, with the additional spin summation in the spinful system.
\item Second, for the case $n=0,m=1$ and $n=1,m=0$, corresponding to the cross terms arising from the bare fermionic excitation operator and the 1st order density-projected fermionic excitation operator,
according to the anticommutation relation of $\hat\psi^{(0)}$ in Eq.\ref{eq:anti_psi0} and $\hat\psi^{(1)}$ as follow
\begin{equation}
\label{eq:anti_psi1_1}
\{\hat{\psi}^{(1)},\,c^\dagger_{i,\alpha,\sigma}\}
=a^{(1)}_{i,\alpha,\sigma}\,n_{i,\alpha,\bar\sigma}-a^{(1)}_{i,\alpha,\bar\sigma}S^{\sigma,\bar\sigma}_{i,\alpha}
\end{equation}
\begin{equation}
\{\hat{\psi}^{(1)\dagger},\,c_{j,\beta,\sigma}\}
=a^{(1)\,*}_{j,\beta,\sigma}\,n_{j,\beta,\bar\sigma}
- a^{(1)\,*}_{j,\beta,\bar\sigma}\,S^{\bar\sigma,\sigma}_{j,\beta}
\label{eq:anti_psi1_2}
\end{equation}
one can obtain
\begin{equation}
\label{eq:chi_01}
\chi_{(i,\alpha);(j,\beta)}^{\mathrm{sub},(0,1)}=\sum_{\sigma} \Big(a^{(0)}_{i,\alpha,\sigma}\,a^{(1)\,*}_{j,\beta,\sigma}\langle n_{j,\beta,\bar\sigma}\rangle_{N-1}-a^{(0)}_{i,\alpha,\sigma}\langle\hat{\psi}^{(1)\dagger}c_{j,\beta,\sigma}\rangle_{N-1}-a^{(1)*}_{j,\beta,\sigma}\langle c_{i,\alpha,\sigma}^\dagger\hat{\psi}^{(0)}n_{j,\beta,\bar\sigma}\rangle_{N-1}+a^{(1)*}_{j,\beta,\bar\sigma}\langle c_{i,\alpha,\sigma}^\dagger\hat{\psi}^{(0)}S_{j,\beta}^{\bar\sigma,\sigma}\rangle_{N-1}\Big)
\end{equation}
\begin{equation}
\label{eq:chi_10}
\chi_{(i,\alpha);(j,\beta)}^{\mathrm{sub},(1,0)}=\sum_{\sigma} \Big(a^{(1)}_{i,\alpha,\sigma}\,a^{(0)\,*}_{j,\beta,\sigma}\langle n_{i,\alpha,\bar\sigma}\rangle_{N-1}-a^{(0)*}_{j,\beta,\sigma}\langle c^\dagger_{i,\alpha,\sigma}\hat{\psi}^{(1)}\rangle_{N-1}
-a^{(1)}_{i,\alpha,\sigma}\langle n_{i,\alpha,\bar\sigma}{\psi}^{(0)\dagger}c_{j,\beta,\sigma}\rangle_{N-1}
+a^{(1)}_{i,\alpha,\bar\sigma}\langle S_{i,\alpha}^{\sigma,\bar\sigma}\hat{\psi}^{(0)\dagger}c_{j,\beta,\sigma}\rangle_{N-1}\Big)
\end{equation}
Here, we use the notation $S_{i,\alpha}^{\sigma,\bar\sigma}\equiv c_{i,\alpha,\sigma}^\dagger c_{i,\beta,\bar\sigma}$, so that $S_{i,\alpha}^{\uparrow,\downarrow}=S_{i,\alpha}^\dagger$ is the spin creation operator and $S_{i,\alpha}^{\downarrow,\uparrow}=S_{i,\alpha}$ is the spin annihilation operator at $(i,\alpha)$. The relations in Eq.\ref{eq:chi_01} and Eq.\ref{eq:chi_10} hold by invoking the total spin conservation of the ground state $\ket{GS_{N-1}}$, so that $\langle S_{i,\alpha}^{\sigma,\bar\sigma}\rangle_{N-1}=\langle S_{j,\beta}^{\bar\sigma,\sigma}\rangle_{N-1}=0$. Moreover, since we have the relation $\sum_{n} \hat{\psi}^{(n)} \ket{GS_{N-1}} = \hat{\psi} \ket{GS_{N-1}} = \hat{\psi}\hat{\psi} \ket{GS_{N}} = 0$, 
the last two terms in Eq.\ref{eq:chi_00} and the second term in Eq.\ref{eq:chi_01} and Eq.\ref{eq:chi_10} can be neglected as a whole when summed over $\chi_{(i,\alpha);(j,\beta)}^{\mathrm{sub},(n,m)}$ for all $n$ and $m$. 

\item Third, for the case $n=m=1$, corresponding to the contribution of the 1st order density-projected fermionic excitation operator to the SDM, according to the anticommutation relations in Eq.\ref{eq:anti_psi1_1} and Eq.\ref{eq:anti_psi1_2}, one can obtain 
\bea
\chi_{(i,\alpha);(j,\beta)}^{\mathrm{sub},(1,1)}
=\sum_{\sigma}\Big[&
a^{(1)}_{i,\alpha,\sigma}a^{(1)\,*}_{j,\beta,\sigma}\;\langle n_{i,\alpha,\bar\sigma}\,n_{j,\beta,\bar\sigma}\rangle_{N-1}\nonumber
-\,a^{(1)}_{i,\alpha,\sigma}a^{(1)\,*}_{j,\beta,\bar\sigma}\;\langle n_{i,\alpha,\bar\sigma}\,S_{j,\beta}^{\bar\sigma,\sigma}\rangle_{N-1}\nonumber
-\,a^{(1)}_{i,\alpha,\bar\sigma}a^{(1)\,*}_{j,\beta,\sigma}\;\langle S_{i,\alpha}^{\sigma,\bar\sigma}\,n_{j,\beta,\bar\sigma}\rangle_{N-1}\nonumber
+\,a^{(1)}_{i,\alpha,\bar\sigma}a^{(1)\,*}_{j,\beta,\bar\sigma}\;\langle S_{i,\alpha}^{\sigma,\bar\sigma}\,S_{j,\beta}^{\bar\sigma,\sigma}\rangle_{N-1}\nonumber\\
&
-\,a^{(1)}_{i,\alpha,\sigma}\;\langle n_{i,\alpha,\bar\sigma}\,\hat\psi^{(1)\dagger} c_{j,\beta,\sigma}\rangle_{N-1}\nonumber
+\,a^{(1)}_{i,\alpha,\bar\sigma}\;\langle S_{i,\alpha}^{\sigma,\bar\sigma}\,\hat\psi^{(1)\dagger} c_{j,\beta,\sigma}\rangle_{N-1}\nonumber
-\,a^{(1)\,*}_{j,\beta,\sigma}\;\langle c_{i,\alpha,\sigma}^\dagger \hat\psi^{(1)}\,n_{j,\beta,\bar\sigma}\rangle_{N-1}\nonumber
+\,a^{(1)\,*}_{j,\beta,\bar\sigma}\;\langle c_{i,\alpha,\sigma}^\dagger \hat\psi^{(1)}\,S_{j,\beta}^{\bar\sigma,\sigma}\rangle_{N-1}\nonumber\Big]
\eea
\bea
\label{eq:chi_11}
\eea
This expression can be further simplified by invoking the total spin-number conservation condition of $\ket{GS_{N-1}}$ such that $\langle n_{i,\alpha,\bar\sigma}\,S_{j,\beta}^{\bar\sigma,\sigma}\rangle_{N-1}=\langle S_{i,\alpha}^{\sigma,\bar\sigma}\,n_{j,\beta,\bar\sigma}\rangle_{N-1}=0$.

\end{itemize}

Taking all above equalities into consideration, one can obtain the overall form of $\rho^{\mathrm{sub}}_{(i,\alpha);(j,\beta)}$ as follow

\begin{align}
\label{eq:final_spinful_SDM}
\rho_{(i,\alpha);(j,\beta)}^{\mathrm{sub}}
=\sum_{\sigma}&\Big[
a^{(0)}_{i,\alpha,\sigma}a^{(0)\,*}_{j,\beta,\sigma}
+a^{(0)}_{i,\alpha,\sigma}a^{(1)\,*}_{j,\beta,\sigma}\;
\big\langle n_{j,\beta,\bar\sigma}\big\rangle_{N-1}
+a^{(1)}_{i,\alpha,\sigma}a^{(0)\,*}_{j,\beta,\sigma}\;
\big\langle n_{i,\alpha,\bar\sigma}\big\rangle_{N-1}
+a^{(1)}_{i,\alpha,\sigma}a^{(1)\,*}_{j,\beta,\sigma}\;
\big\langle n_{i,\alpha,\bar\sigma}\,n_{j,\beta,\bar\sigma}\big\rangle_{N-1}\nonumber
+\,a^{(1)}_{i,\alpha,\bar\sigma}a^{(1)\,*}_{j,\beta,\bar\sigma}\;
\big\langle S_{i,\alpha}^{\sigma,\bar\sigma}\,S_{j,\beta}^{\bar\sigma,\sigma}\big\rangle_{N-1}\nonumber
\\
&
+
\,a^{(1)}_{i,\alpha,\bar\sigma}\;\langle S_{i,\alpha}^{\sigma,\bar\sigma}\,\hat\psi^{\dagger} c_{j,\beta,\sigma}\rangle_{N-1}\nonumber
+\,a^{(1)\,*}_{j,\beta,\bar\sigma}\;\langle c_{i,\alpha,\sigma}^\dagger \hat\psi\,S_{j,\beta}^{\bar\sigma,\sigma}\rangle_{N-1}\nonumber
-\,a^{(1)}_{i,\alpha,\sigma}\;\langle n_{i,\alpha,\bar\sigma}\,\hat\psi^{\dagger} c_{j,\beta,\sigma}\rangle_{N-1}\nonumber
-\,a^{(1)\,*}_{j,\beta,\sigma}\;\langle c_{i,\alpha,\sigma}^\dagger \hat\psi\,n_{j,\beta,\bar\sigma}\rangle_{N-1}\nonumber
\\&+\langle  c_{i,\alpha,\sigma}^\dagger \hat\psi^\dagger\hat\psi c_{j,\beta,\sigma}\rangle_{N-1}+\cdots\Big]
\end{align}
The second and third lines of this expression can be further expanded order by order and simplified using the following commutation relations:
\begin{equation}
[\hat\psi,n_{j,\beta,\bar\sigma}]=a^{(0)}_{j,\beta,\bar\sigma}c_{j,\beta,\bar\sigma}+a^{(1)}_{j,\beta,\bar\sigma}c_{j,\beta,\bar\sigma}n_{j,\beta,\sigma}+\cdots
\end{equation}
\begin{equation}
[n_{i,\alpha,\bar\sigma},\hat\psi^\dagger]=a^{(0)*}_{i,\alpha,\bar\sigma}c^\dagger_{i,\alpha,\bar\sigma}+a^{(1)*}_{i,\alpha,\bar\sigma}c^\dagger_{i,\alpha,\bar\sigma}n_{i,\alpha,\sigma}+\cdots
\end{equation}
\begin{equation}
[\hat\psi,S^{\bar\sigma,\sigma}_{j,\beta}]=a^{(0)}_{j,\beta,\bar\sigma}c_{j,\beta,\sigma}+a^{(1)}_{j,\beta,\bar\sigma}c_{j,\beta,\sigma}n_{j,\beta,\bar\sigma}+\cdots
\end{equation}
\begin{equation}
[S^{\sigma,\bar\sigma}_{i,\alpha},\hat\psi^\dagger]=a^{(0)*}_{i,\alpha,\bar\sigma}c^\dagger_{i,\alpha,\sigma}+a^{(1)*}_{i,\alpha,\bar\sigma}c^\dagger_{i,\alpha,\sigma}n_{i,\alpha,\bar\sigma}+\cdots
\end{equation}
\begin{equation}
\{\hat\psi,c_{j,\beta,\sigma}\}=-a^{(1)}_{j,\beta,\bar\sigma}c_{j,\beta,\bar\sigma}c_{j,\beta,\sigma}+\cdots
\end{equation}
\begin{equation}
\{c_{i,\alpha,\sigma}^\dagger,\hat\psi^\dagger\}=-a^{(1)*}_{i,\alpha,\bar\sigma}c^\dagger_{i,\alpha,\sigma}c^\dagger_{i,\alpha,\bar\sigma}+\cdots
\end{equation}
Combining this with the fact that $\hat\psi \ket{GS_{N-1}} = \hat\psi \hat\psi \ket{GS_{N}} = 0$ and the total spin quantum number conservation, Eq.~\ref{eq:final_spinful_SDM} can be rewritten in its final, order-by-order simplified form as 
\bea
\rho_{(i,\alpha);(j,\beta)}^{\mathrm{sub}}
=\sum_{\sigma}&\Big[
a^{(0)}_{i,\alpha,\sigma}a^{(0)\,*}_{j,\beta,\sigma}
+a^{(0)}_{i,\alpha,\sigma}a^{(1)\,*}_{j,\beta,\sigma}\;
\big\langle n_{j,\beta,\bar\sigma}\big\rangle_{N-1}
+a^{(1)}_{i,\alpha,\sigma}a^{(0)\,*}_{j,\beta,\sigma}\;
\big\langle n_{i,\alpha,\bar\sigma}\big\rangle_{N-1}
+a^{(1)}_{i,\alpha,\sigma}a^{(1)\,*}_{j,\beta,\sigma}\;
\big\langle n_{i,\alpha,\bar\sigma}\,n_{j,\beta,\bar\sigma}\big\rangle_{N-1}\nonumber
+\,a^{(1)}_{i,\alpha,\bar\sigma}a^{(1)\,*}_{j,\beta,\bar\sigma}\;
\big\langle S_{i,\alpha}^{\sigma,\bar\sigma}\,S_{j,\beta}^{\bar\sigma,\sigma}\big\rangle_{N-1}\nonumber
\\
&
+(a_{i,\alpha,\bar\sigma}^{(1)}a_{i,\alpha,\bar\sigma}^{(0)*}+a_{j,\beta,\bar\sigma}^{(1)}a_{j,\beta,\bar\sigma}^{(0)*})\langle c_{i,\alpha,\sigma}^\dagger c_{j,\beta,\sigma}\rangle_{N-1} + |a_{i,\alpha,\bar\sigma}^{(1)}|^2\langle c_{i,\alpha,\sigma}^\dagger n_{i,\alpha,\bar\sigma}c_{j,\beta,\sigma}\rangle_{N-1}+ |a_{j,\beta,\bar\sigma}^{(1)}|^2\langle c_{i,\alpha,\sigma}^\dagger n_{j,\beta,\bar\sigma} c_{j,\beta,\sigma}\rangle_{N-1}\\
&
+a^{(1)*}_{i,\alpha,\bar\sigma}a^{(1)*}_{j,\beta,\bar\sigma}\langle c^\dagger_{i,\alpha,\sigma}c^\dagger_{i,\alpha,\bar\sigma}c_{j,\beta,\bar\sigma}c_{j,\beta,\sigma}\rangle_{N-1}+\cdots\Big]
\eea
\begin{equation}
\label{eq:simplified_order_by_order_spinful_SDM}
\end{equation}

%\textcolor{blue}{Is this argument proper: }
In summary, the zeroth-order contribution $\rho^{\mathrm{sub},(0)}$ corresponds directly to the MDM 
of the bare single-particle excitation $\psi^{(0)}$, fully consistent with the spinless case. 
%For spinful systems
Nevertheless, additional corrections arise from the projected fermionic excitation $\psi^{(1)}$, 
involving density average $\big\langle n_{i,\alpha,\bar\sigma}\big\rangle_{N-1}$, density--density correlators $\big\langle n_{i,\alpha,\bar\sigma}\,n_{j,\beta,\bar\sigma}\big\rangle_{N-1}$
and spin--spin correlators $\big\langle S_{i,\alpha}^{\sigma,\bar\sigma}\,S_{j,\beta}^{\bar\sigma,\sigma}\big\rangle_{N-1}$.
We expect density-density and spin-spin correlations in systems without magnetic or charge order to decay rapidly with oscillations and thus be strongly suppressed at large separations $x$, while the density average contributes a subleading, exponentially decaying modification $\Big(a^{(0)}_{i,\alpha,\sigma}a^{(1)\,*}_{j,\beta,\sigma}\;
\big\langle n_{j,\beta,\bar\sigma}\big\rangle_{N-1}
+a^{(1)}_{i,\alpha,\sigma}a^{(0)\,*}_{j,\beta,\sigma}\;
\big\langle n_{i,\alpha,\bar\sigma}\big\rangle_{N-1}\Big)$ at finite $x$.
For the terms in the second and third lines, involving the structures of single-particle correlators $\langle c_{i,\alpha,\sigma}^\dagger c_{j,\beta,\sigma}\rangle_{N-1}$, density-projected single-particle correlators $\langle c_{i,\alpha,\sigma}^\dagger n_{i,\alpha,\bar\sigma}c_{j,\beta,\sigma}\rangle_{N-1}$ and pair-pair correlators $\langle c^\dagger_{i,\alpha,\sigma}c^\dagger_{i,\alpha,\bar\sigma}c_{j,\beta,\bar\sigma}c_{j,\beta,\sigma}\rangle_{N-1}$.
We argue that these contributions are small: for weak interactions (small $U$), the coefficients $|a^{(1)}_{i,\alpha,\sigma}|$ are negligible, 
while for strong interactions (large $U$) the double occupancy is strongly suppressed. Combining with the fact that single-particle correlators and pair-pair correlators also decay rapidly with oscillatory behavior, 
the expectation values of the terms in the second and third lines are driven toward zero. 
Taken together, we expect that $\rho^{\mathrm{sub},(0)}$ still provides the main contribution to the SDM, retaining its clean exponential decay even at finite interaction $U$ in the Anderson--Hubbard model, as shown in Fig.\ref{fig:2d_correlation}(b) and Fig.\ref{fig:supp_2d_corr_fit}.
In more general situations, we cannot prove this feature is always true. However, as we discussed in the main text,
we expect the ground states to be always localized for finite-size quasi-1D systems at finite disorder strength except in special cases.
If there exists one $\hat{\psi}$, SDM still provides important information about $\hat{\psi}$.

\section{Details of DMRG calculations for interacting systems}

In this section we provide additional details regarding the numerical implementation of our method in interacting disordered systems. 
We first describe the DMRG parameters used in the calculations of the spinless 1D interacting model (Eq.\ref{eq:attractive_model}) and the Anderson--Hubbard model (Eq.\ref{eq:anderson-hubbard}) in quasi-1D system with width $M=2,3$, together with benchmark comparisons to ED at $V=0$ ($U=0$) for randomly selected disorder samples. 
We also present supplementary results for the quasi-1D Anderson--Hubbard model, including $\gamma(x)$ obtained at different disorder strengths $W$ and interaction strengths $U$. 
These results complement the discussion in the main text and further confirm the accuracy and robustness of the many-body MDM approach.

\subsection{DMRG parameters and benchmarks}
\label{supp:benchmark}
In this work, numerical calculations for interacting systems are performed using the DMRG method \cite{DMRG_PhysRevLett.69.2863}, as implemented in the ITensor package \cite{DMRG_Itensor_1,DMRG_Itensor_2}.
 We first describe the DMRG setup used for the spinless 1D interacting model with onsite disorder. 
In our calculation, we consider the systems of size $L=120$ at half-filling in the spinless case. We exploited particle-number conservation, restricting the system to particle numbers $N=60$ and $N=59$ in order to compute the two corresponding ground states $\ket{GS_N}$ and $\ket{GS_{N-1}}$.  
The number of kept states is sequentially increased following the array $[30,50,100,200,400]$, in combination with finite-lattice sweeps given by $[80,80,80,80,10]$.
From the two ground states obtained in this way, the SDM is evaluated following Eq.\ref{eq:SDM_definition}, and the interacting version of the MDM is then constructed according to Eq.\ref{eq:SDM_MDM}. 
After symmetrization, the slowest-decaying mode $\gamma(x)$ is extracted from the MDM.

\begin{figure}[H]
    \hspace*{-0.8cm} 
    \includegraphics[width=1.08\linewidth]{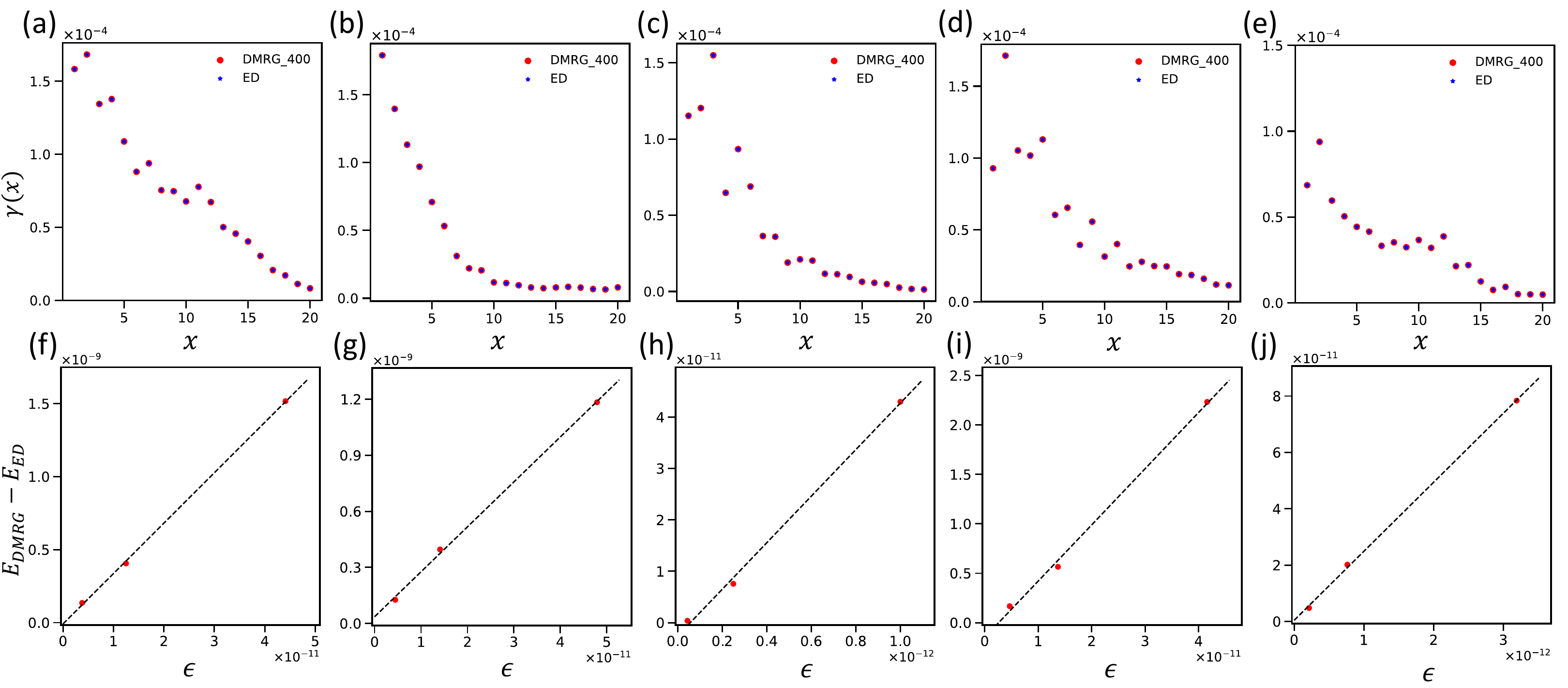}
    \caption{ Benchmarks results between DMRG and ED for the spinless 1D model with onsite disorder (Eq.\ref{eq:attractive_model}) of size $L=120$ at $U=0$, $W=2t$ and half-filling under OBC.  
(a)--(e) show $\gamma(x)$ as a function of $x$ for five randomly chosen disorder samples.  
Red dots are DMRG results with the number of kept states gradually increased up to 400, while blue dots are ED results.  
(f)-–(j) show the scaling of the ground-state energy difference $E_{\mathrm{DMRG}} - E_{\mathrm{ED}}$ for $\ket{GS_N}$ with truncation error $\epsilon$ for the corresponding samples.  
When performing the site average, the 20 sites closest to each boundary are excluded to minimize open-boundary effects.}
    \label{fig:bench_L1}
\end{figure}

In the case of $V=0$, the MDM obtained from DMRG agrees extremely well with that from ED of the non-interacting Hamiltonian. 
To benchmark the DMRG results against ED at the single-sample level, the SDM is averaged over lattice sites only, without averaging over disorder realizations.
Five disorder realizations are randomly chosen for comparison.
Fig.\ref{fig:bench_L1}(a)–(e) display $\gamma(x)$ obtained from both DMRG (red dots) and non-interacting ED (blue dots), showing excellent agreement.
Fig.~\ref{fig:bench_L1}(f)–(j) further present the scaling of the ground-state energy difference $\Delta E=E_{\mathrm{DMRG}}-E_{\mathrm{ED}}$ for $\ket{GS_N}$ between DMRG and ED as a function of truncation error $\epsilon$. All five samples reach $\epsilon < 10^{-10}$ with $\Delta E < 10^{-8}$.
These results demonstrate that our DMRG setup converges to the correct ground states in this model and validate the reliability of the calculations employed in the interacting case.

\begin{figure}[H]
    \hspace*{-0.8cm} 
    \includegraphics[width=1.08\linewidth]{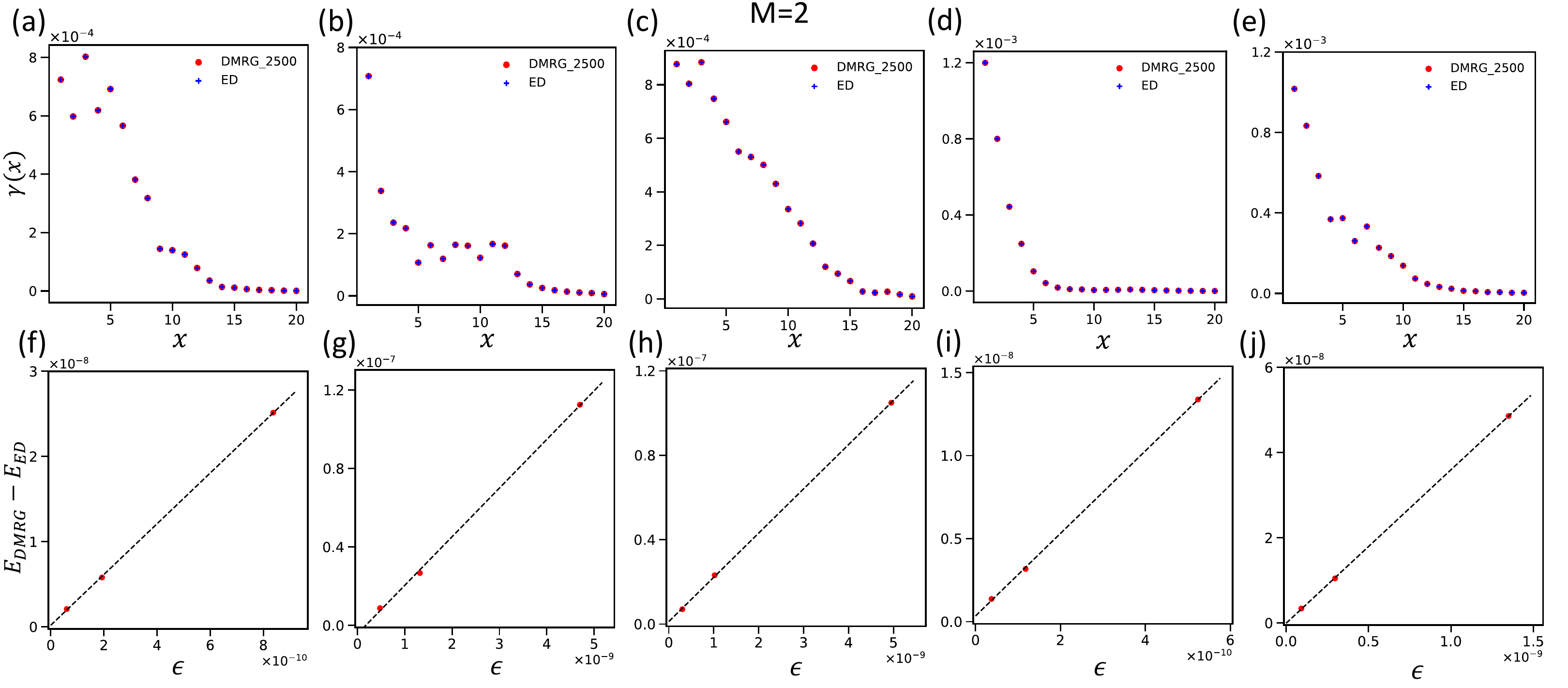}
    \caption{ Benchmarks for the quasi-1D Anderson--Hubbard (Eq.\ref{eq:anderson-hubbard}) model of size $2\times 60$ at $U=0$, $W=2.5t$ and filling $n=4/15$ under OBC.  
(a)--(e) show $\gamma(x)$ as a function of $x$ for five randomly chosen disorder samples.  
Red dots are DMRG results with the number of kept states gradually increased up to 2500, while blue dots are ED results.  
(f)–-(j) show the scaling of the ground-state energy difference $E_{\mathrm{DMRG}} - E_{\mathrm{ED}}$ for $\ket{GS_N}$ with truncation error $\epsilon$ for the corresponding samples.  
When performing the site average, the 10 sites closest to each boundary are excluded to minimize open-boundary effects.
}
    \label{fig:bench_L2}
\end{figure}

We next introduce the DMRG setup for the quasi-1D Anderson--Hubbard model and present benchmark comparisons with non-interacting ED at $U=0$ for five randomly chosen disorder realizations. 
For the $M=2$ system, we consider system with size $2\times 60$ at filling $n=4/15$, enforcing both particle-number and spin conservation with $N_\uparrow=32$, $N_\downarrow=32$ for $\ket{GS_N}$, and $N_\uparrow=31$, $N_\downarrow=32$ for $\ket{GS_{N-1}}$. 
The number of kept states is increased sequentially following the array $[50,100,200,400,800,1000,1500,2000,2500]$ with finite-lattice sweeps $[25,25,25,25,25,25,5,5,5]$. 
Five randomly chosen disorder samples are computed by both DMRG and non-interacting ED at $U=0$. 
Figures.\ref{fig:bench_L2}(a)–(e) show $\gamma(x)$ obtained from the two methods, exhibiting perfect agreement. 
Figures.\ref{fig:bench_L2}(f)–(j) further display the scaling of the ground-state energy difference $\Delta E=E_{\mathrm{DMRG}}-E_{\mathrm{ED}}$ with truncation error $\epsilon$, which exhibits an almost linear relation with intercepts close to zero. 
The final truncation error for all samples is below $10^{-8}$. 

\begin{figure}[H]
    \hspace*{-0.8cm} 
    \includegraphics[width=1.08\linewidth]{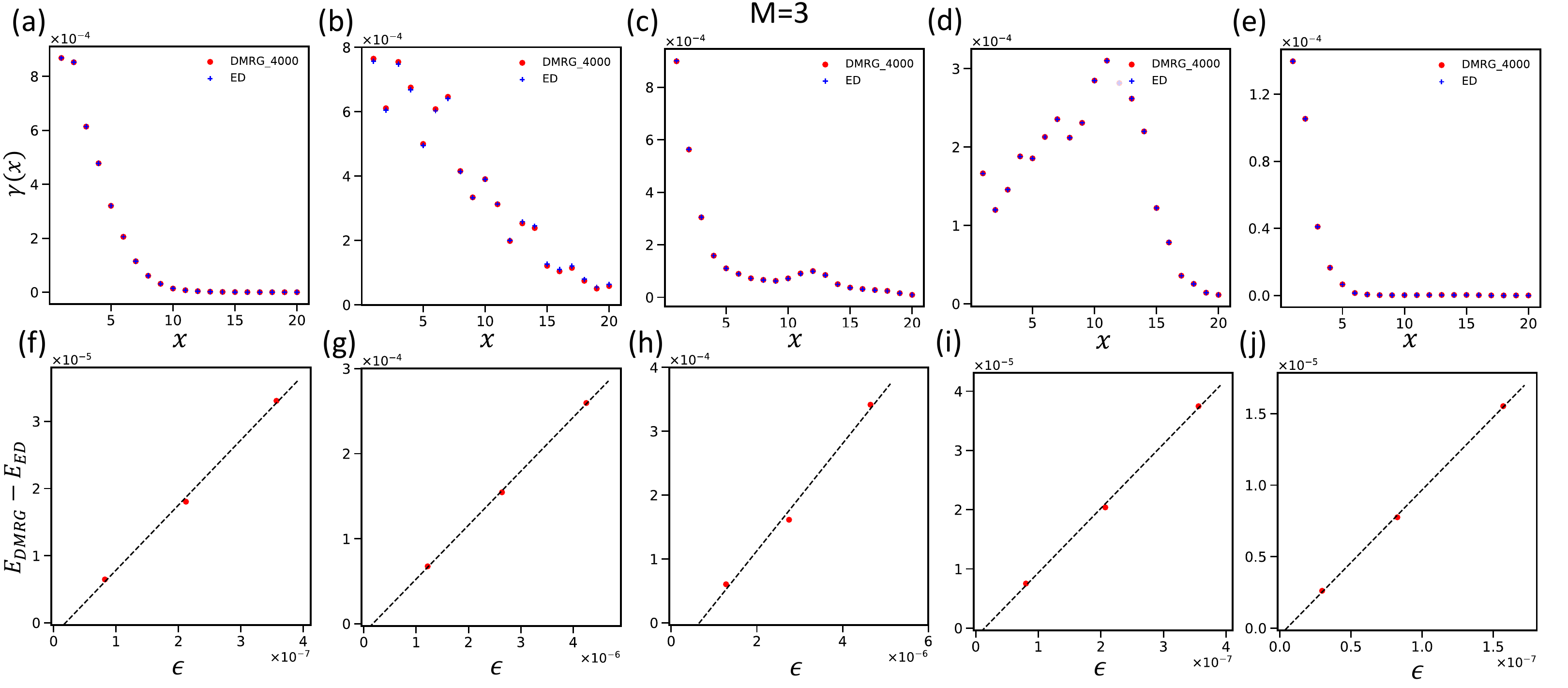}
    \caption{ Benchmarks for the quasi-1D Anderson--Hubbard (Eq.\ref{eq:anderson-hubbard}) model of size $3\times 60$ at $U=0$, $W=2.5t$ and filling $n=4/15$ under OBC.  
(a)--(e) show $\gamma(x)$ as a function of $x$ for five randomly chosen disorder samples.  
Red dots are DMRG results with the number of kept states gradually increased up to 4000,  while blue dots are ED results.  
(f)–-(j) show the scaling of the ground-state energy difference $E_{\mathrm{DMRG}} - E_{\mathrm{ED}}$ for $\ket{GS_N}$ with truncation error $\epsilon$ for the corresponding samples.  
When performing the site average, the 10 sites closest to each boundary are excluded to minimize open-boundary effects.
}
    \label{fig:bench_L3}
\end{figure}

For the $M=3$ system of size $3\times 60$ at the same filling $n=4/15$, we perform calculations with $N_\uparrow=48$, $N_\downarrow=48$ for $\ket{GS_N}$ and $N_\uparrow=47$, $N_\downarrow=48$ for $\ket{GS_{N-1}}$.
Here the number of kept states is increased following the array $[50,100,200,400,800,1000,1500,2000,2500,3000,4000]$ with finite-lattice sweeps $[25,25,25,25,25,25,15,15,5,5,5]$. 
Again, five disorder realizations are randomly chosen for comparison between DMRG and noninteracting ED at $U=0$.
The $\gamma(x)$ results in Fig.\ref{fig:bench_L3}(a)-(e) show excellent agreement between the two methods, while the energy differences $\Delta E$ scale linearly with truncation error $\epsilon$ and extrapolate to zero, with final truncation errors below $10^{-6}$, as shown in Fig.\ref{fig:bench_L3}.  
These benchmarks confirm that our DMRG setup for the quasi-1D systems of width $M=2$ and $M=3$ yields well-converged ground states and provides reliable results for subsequent MDM analysis.

\subsection{Supplementary results for the Anderson--Hubbard model}
\label{supp:anderson-hubbard}
\begin{figure}[H]
    \raggedleft
    \hspace*{-1.5cm} 
    \includegraphics[width=1.15\linewidth]{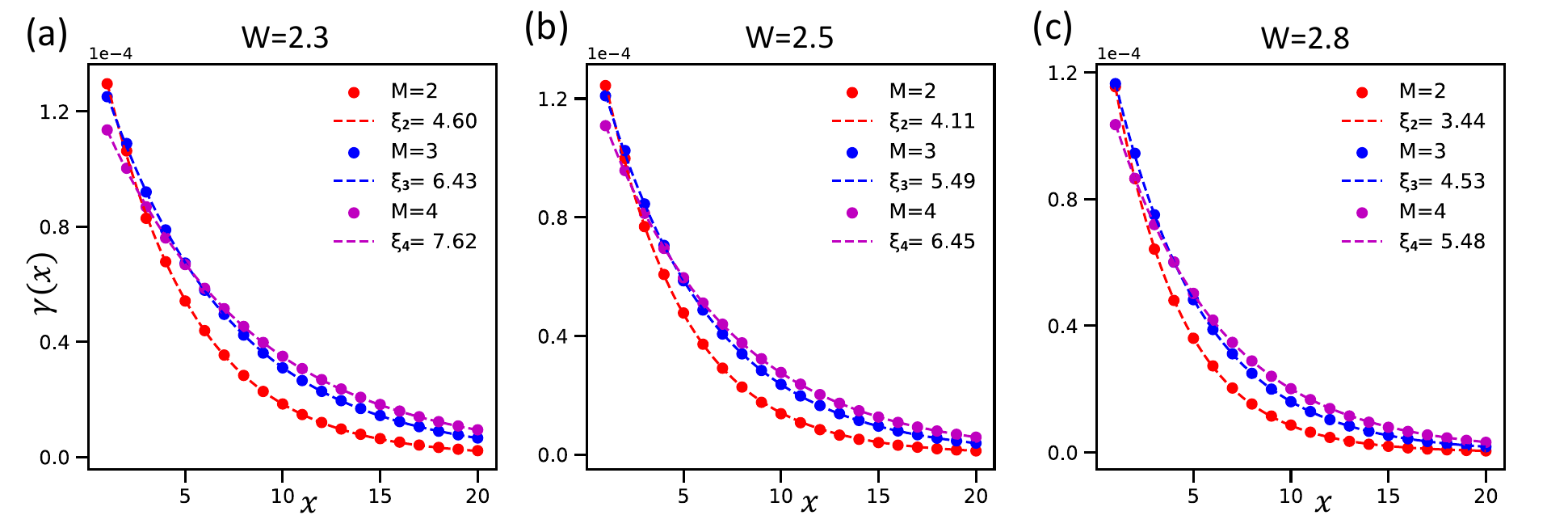}
    \caption{ Results of $\gamma(x)$ for the quasi-1D Anderson-Hubbard model at $U=0t$ and filling $n=4/15$. (a)--(c) show the $\gamma(x)$ together with exponential fits results for $M=2,3,4$ in the non-interacting case with disorder strengths $W=2.3t, 2.5t, 2.8t$, respectively.  These results are calculated by ED of non-interacting Hamiltonian. All disorder realizations are on systems of size $M\times L$ with $L=60$ under OBC, and in the site averaging for each sample, the 10 sites nearest to each boundary are excluded to minimize open-boundary effects. Each curve is calculated by averaging 8000 disorder realizations.   }
    \label{fig:supp_2d_nonint_fit}
\end{figure}

\begin{figure}[H]
    \hspace*{-0.9cm} 
    \includegraphics[width=1.12\linewidth]{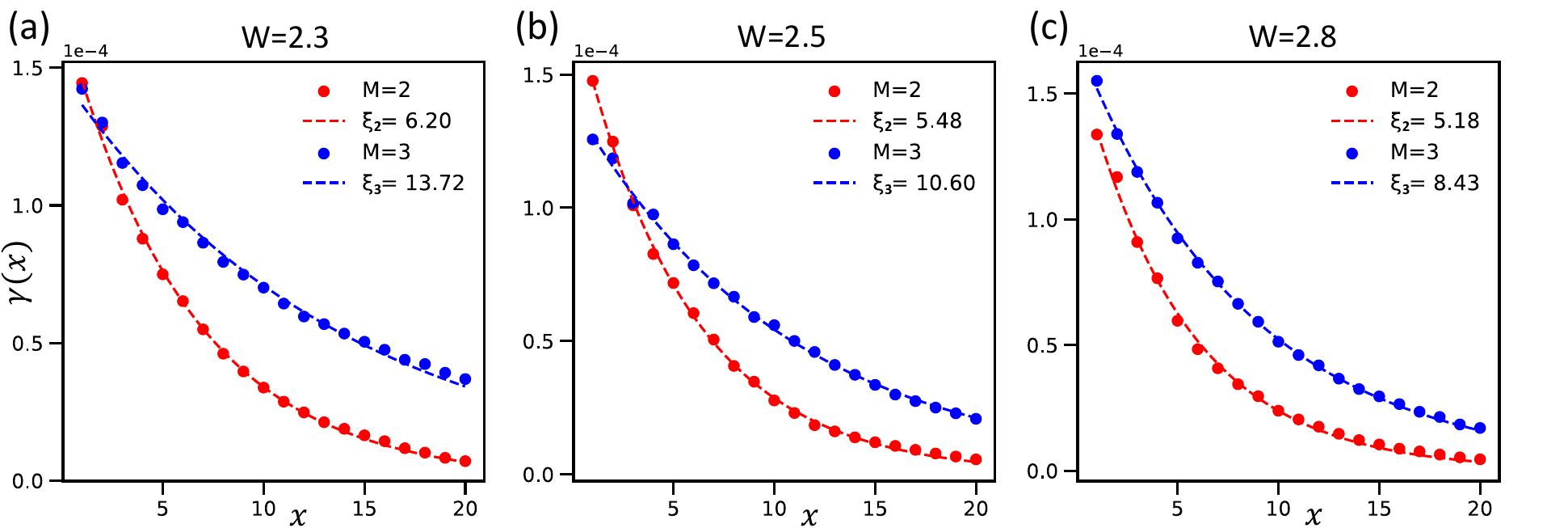}
    \caption{ Results of $\gamma(x)$ for the quasi-1D Anderson-Hubbard model at $U=4t$ and filling $n=4/15$. (a)--(c) show the $\gamma(x)$ together with exponential fits results for $M=2,3,4$ with disorder strengths $W=2.3t, 2.5t, 2.8t$, respectively.  These results are calculated by DMRG. All disorder realizations are on systems of size $M\times L$ with $L=60$ under OBC, and in the site averaging for each sample, the 10 sites nearest to each boundary are excluded to minimize open-boundary effects. Each curve is calculated by averaging 100 disorder realizations.}
    \label{fig:supp_2d_corr_fit}
\end{figure}

We now present additional results for the quasi-1D Anderson--Hubbard model, focusing on the $\gamma(x)$ results corresponding to the finite-size scaling data corresponding to the finite-size scaling shown in Figs.~\ref{fig:2d_correlation}(c), (d). 
For the non-interacting case ($U=0$), Fig.\ref{fig:supp_2d_nonint_fit} shows $\gamma(x)$ at disorder strengths $W=2.3t,2.5t,2.8t$ for bars of width $M=2,3,4$. 
In all cases, $\gamma(x)$ exhibits a clear exponential decay, $\gamma(x)\sim e^{-x/\xi_M}$, from which localization lengths $\xi_M$ can be directly extracted. 
Remarkably, even in the presence of strong Hubbard interactions, $\gamma(x)$ retains a simple exponential form, as shown in Fig.~\ref{fig:supp_2d_corr_fit} for $M=2,3$ at $U=4t$ and the same disorder strengths.
Moreover, for fixed $M$ and $W$, the localization length at $U=4t$ is significantly enhanced compared to the $U=0$ case, demonstrating the delocalizing effect of Hubbard interactions in disordered systems. 
This enhancement increases with $M$, ultimately leading to the emergence of a correlated metallic state at finite $U$ and $W$.

\end{document}